\documentclass[11pt,a4paper]{article}
\usepackage{jcappub}

\usepackage{amsmath,amssymb,bm,float,mathrsfs,subfigure,tabularx}
\usepackage{graphicx}
\usepackage{dcolumn}
\usepackage{color}
\usepackage{hyperref}

  \def\vs#1{\vspace{#1\baselineskip}}
  
  \newcommand{\beq}{\begin{equation}}
  \newcommand{\eeq}{\end{equation}}
  \newcommand{\al}[1]{\begin{align} #1 \end{align}}
  \newcommand{\bi}{\begin{itemize}}
  \newcommand{\ei}{\end{itemize}}
  \newcommand{\bc}{\begin{center}}
  \newcommand{\ec}{\end{center}}
  
  \def\dd{\mathrm{d}}
  \def\DD{\mathrm{D}}
  \def\rmg{\mathrm{g}}
  \def\rmm{\mathrm{m}}
  \def\rmB{\mathrm{B}}
  \def\rmD{\mathrm{D}}
  \def\rmE{\mathrm{E}}

  \def\rmS{\mathrm{S}}

  \def\rmX{\mathrm{X}}

  \def\mcD{\mathcal{D}}

  \def\mcG{\mathcal{G}}

  \def\mcK{\mathcal{K}}
  
  \def\mcM{\mathcal{M}}
  
  \def\mcO{\mathcal{O}}
  \def\mcP{\mathcal{P}}
  
  \def\mcR{\mathcal{R}}
  \def\mcS{\mathcal{S}}
  \def\mcT{\mathcal{T}}
  
  \def\mcV{\mathcal{V}}

  \def\pd{\partial}

  \def\pspin{\ooalign{\hfil/\hfil\crcr$\partial$}}
  \def\mspin{\bar{\ooalign{\hfil/\hfil\crcr$\partial$}}}
  \newcommand{\ave}[1]{\left\langle #1 \right\rangle}

  \def\3Dint#1{\int\frac{\dd^{3}{#1 }}{(2\pi )^3}}

  \def\hatn{\hat{\bm n}}

  \def\cS{\chi_{\rm S}}

{\baselineskip0pt
\rightline{\baselineskip16pt\rm\vbox to20pt{
            \hbox{RESCEU-6/13, YITP-13-35}
\vss}}%
}

\title{Full-sky formulae for weak lensing power spectra from total 
angular momentum method}

\author[a]{Daisuke Yamauchi}
\author[b]{Toshiya Namikawa}
\author[a,c]{Atsushi Taruya}

\affiliation[a]{%
Research Center for the Early Universe, Graduate School of Science, 
The University of Tokyo, Bunkyo-ku, Tokyo 113-0033, Japan
}%
\affiliation[b]{%
Yukawa Institute for Theoretical Physics, Kyoto University, Kyoto 606-8502, Japan
}%
\affiliation[c]{%
Kavli Institute for the Physics and Mathematics of the Universe, The University 
of Tokyo, Kashiwa, Chiba 277-8568, Japan 
}%

\emailAdd{yamauchi@resceu.s.u-tokyo.ac.jp}
\emailAdd{namikawa@yukawa.kyoto-u.ac.jp}
\emailAdd{ataruya@utap.phys.s.u-tokyo.ac.jp}

\abstract{
We systematically derive full-sky formulae for the weak lensing power 
spectra generated by scalar, vector and tensor perturbations 
from the total angular 
momentum (TAM) method.  Based on both the geodesic and geodesic deviation 
equations, 
we first give the gauge-invariant expressions for the deflection angle and 
Jacobi map as observables of the CMB lensing and cosmic shear experiments. 
We then apply the TAM method, originally 
developed in the theoretical studies of CMB, to a systematic derivation of
the angular power spectra. The TAM representation, which characterizes the total angular dependence of the spatial modes projected along a line-of-sight, 
can carry all the information of the lensing modes 
generated by scalar, vector, and tensor metric perturbations. This greatly 
simplifies the calculation, and we present 
a complete set of the full-sky formulae for angular power spectra in both 
the E-/B-mode cosmic shear and gradient-/curl-mode lensing potential of 
deflection angle. 
Based on the formulae, we give illustrative examples of non-vanishing 
B-mode cosmic shear and curl-mode of deflection angle 
in the presence of the vector and tensor perturbations, and explicitly 
compute the power spectra. 
}

\begin{document}

\maketitle
\tableofcontents

\section{Introduction}
\label{sec:Introduction}

Precision weak lensing measurement in cosmology is the key to improve our 
view of the Universe, and it directly offers an opportunity to probe 
unseen cosmological fluctuations along a line-of-sight of photon path.   
In particular, planned wide- and deep-imaging surveys such as 
Subaru Hyper Supreme-Can (HSC)~\cite{HSC}, Dark Energy Survey 
(DES)~\cite{astro-ph/0510346}, 
Euclid~\cite{Refregier:2010ss}, and Large Synaptic Survey Telescope 
(LSST)~\cite{arXiv:0912.0201} will provide a high-precision measurement of 
the deformation of the distant-galaxy images, whose non-vanishing spatial 
correlation is primarily caused by the gravitational lensing. 
The so-called cosmic shear is now recognized as a 
standard cosmological tool, and plays an important role to 
constrain the growth of structure and/or cosmic expansion \cite{Pitrou:2012ge,Bernardeau:2011tc,Bernardeau:2009bm,Seitz:1994xf,Kaiser:1996tp,Blandford:1991zz,Sasaki:1987ad,1961RSPSA.264..309S,Hu:2000ee,Stebbins:1996wx,Kamionkowski:1997mp} 
(for reviews, see \cite{Lewis:2006fu,Perlick:2004tq,Bartelmann:1999yn,Sasaki:1993tu}). On the other 
hand, cosmic microwave background also carries the information on the 
gravitational lensing, and through the sophisticated reconstruction 
techniques, one can measure the gravitational lensing deflection of the 
CMB photons, referred to as the CMB-lensing signals. The ground-based 
experiments,  
Atacama Cosmology Telescope (ACT)~\cite{Das:2011ak,Das:2013zf}
and South Pole Telescope (SPT)~\cite{vanEngelen:2012va}, 
as well as Planck satellite have already revealed the 
undoubted lensing signals~\cite{Ade:2013mta}, and future CMB experiments,
including, POLARBEAR~\cite{arXiv:1011.0763}, 
ACTPol~\cite{arXiv:1006.5049}, SPTPol~\cite{2009AIPC.1185..511M}, 
CMBPol~\cite{Baumann:2008aq}, and COrE~\cite{Bouchet:2011ck},  
will measure the lensing deflection field much more precisely. 

While the weak lensing effect detected and measured so far is mostly
dominated by the scalar metric perturbation induced by the 
large-scale structure, with an increased precision, a search for tiny 
signals generated by the vector and tensor perturbations is made possible, 
and the detection and/or measurement of such signals would provide 
a valuable insight into the physics and history of the very early universe. 
Theoretically, 
the distortion effect of lensing on the primary CMB anisotropies is expressed by a remapping with 
two dimensional vector, usually referred to as deflection angle, which can be estimated through 
the fact that a fixed lensing potential introduces statistical anisotropy into the observed CMB. 
Hence we consider the CMB lensing as being a solution of geodesic equation.
On the other hand, for galaxy survey, what we can measure is the shape (or shear) 
of galaxies modified by gravitational lensing, which is characterized 
by the deformation of two-dimensional spatial pattern. 
Therefore we solve the geodesic deviation equation for the shear field.
The lensing fields can be generally decomposed into two modes: 
the gradient of scalar lensing potential (gradient-mode)
and rotation of pseudo-scalar lensing potential (curl-mode) for 
deflection angle (e.g., \cite{Stebbins:1996wx,Hirata:2003ka,Namikawa:2011cs}), and 
the even- and odd-parity modes (E-/B-modes) for cosmic shear 
(e.g., \cite{Stebbins:1996wx,Kamionkowski:1997mp}). 
One important aspect from the symmetric argument is that 
the scalar perturbation can produce both the E-mode shear and 
the gradient-mode lensing potential of the deflection angle, 
while it is unable to generate the B-mode shear 
and the curl-mode lensing potential. Hence, non-vanishing B-mode or 
curl-mode signal on large angular scales immediately implies the presence of 
non-scalar metric perturbations.

In this paper, we systematically derive the angular power spectra of 
gradient-/curl-mode lensing potential and E-/B-mode cosmic shear, and 
present a complete set of full-sky formula for scalar, vector, and tensor
metric perturbations. As illustrative examples, 
we consider the cosmic strings and primordial gravitational waves as 
representative sources for vector and tensor perturbations. 
Based on the formulae, we explicitly compute the power spectra, showing 
that the non-vanishing B-mode and curl-mode lensing signals 
naturally arise. 
In deriving the weak lensing power spectra, one complication is that 
while the weak lensing observables are defined on a spherical sky, 
the metric perturbations as the source of the gravitational lensing usually 
appear in the three-dimensional space. Thus, even decomposing the 
perturbations into the plane waves, due to their angular structure, 
a plane wave along a line-of-sight can contribute to the lensing 
observables at several multipoles. The situation is more 
complicated for the vector and tensor 
perturbations, since the helicity basis of their perturbations also has 
explicit angular dependence, and contributes to many multipoles. 
One way to avoid these complications is to isolate the total angular 
dependence of the perturbations by introducing new representation, 
the {\it total angular momentum (TAM) representation}~\cite{Hu:1997hp} 
(see also \cite{Dai:2012bc}), originally 
developed in the theoretical studies of CMB. 
Combining the intrinsic angular structure 
with that of the plane-wave spatial dependence, 
the total angular momentum representation substantially simplifies the 
derivation of the full-sky formula, and it enables us to simultaneously 
treat the weak lensing by vector and tensor perturbations on an equal 
footing with those generated by scalar perturbation. 
As a result, we obtain a complete set of power spectra in both 
the cosmic shear and lensing potential of deflection angle. 
Our full-sky formulae rigorously coincide with those obtained previously
based on a more involved calculation
(see \cite{Yamauchi:2012bc,Dodelson:2003bv,Schmidt:2012ne,Cooray:2005hm,Li:2006si,Schmidt:2012nw,Sarkar:2008ii}
for vector and tensor perturbations).


The paper is organized as follows.
We begin by defining the unperturbed and perturbed spacetime metrics, 
and quantities associated with the photon path 
in section \ref{sec:Background and perturbations}. 
In section \ref{sec:Weak lensing observables}, based on the 
gauge-invariant formalism, 
we solve the geodesic and geodesic deviation equations in
the presence of all types of the metric perturbations. 
We then derive the explicit relation between the deflection angle and 
the deformation matrix. Section \ref{sec:Total angular momentum method} is 
the main part of this paper.  We introduce the TAM 
representation, and rewrite the expression of the lensing observables 
in term of this representation. Making full use of the properties of 
the 
TAM representation, 
we present the full-sky formulae for angular power spectra of 
the E-/B-mode cosmic 
shear and the gradient-/curl-mode lensing potential. 
In section \ref{sec:Applications}, based on the formulae, 
we give illustrative examples of non-vanishing 
B-mode and curl-mode lensing signals
in the presence of the vector and tensor perturbations, and explicitly 
compute the power spectra. 
Finally, section \ref{sec:Summary} is devoted to summary and conclusion.
Throughout the paper, we assume a flat $\Lambda$CDM cosmological model
with the cosmological parameters : $\Omega_{\rm b} h^2=0.22$\,, 
$\Omega_\rmm h^2=0.13$\,, $\Omega_\Lambda =0.72$\,, $h=0.7$\,,
$n_{\rm s} =0.96$\,, $\Delta_\mcR^2 (k_{\rm pivot}) =2.4\times 10^{-9}$\,,
$k_{\rm pivot}=0.002{\rm Mpc}^{-1}$\,, $r=0.2$\,, 
$n_{\rm t}=-r/8=-0.025$\,, $\tau =0.086$\,.
In Table \ref{notation}, we list the definition 
of the quantities used in the paper. 

\begin{table}[t]
\caption{
Notations for quantities used in this paper.
}
\vs{0.5}
\begin{tabular}{ccc} \hline \hline 
Symbol & eq. & Definition \\ 
\hline 
$g_{\mu\nu}$&\eqref{eq:unperturbed 4-dimensional metric}
& $4$-dimensional metric on unperturbed spacetime\\
$\bar\gamma_{ij}$&\eqref{eq:unperturbed 4-dimensional metric}
&$3$-dimensional spatial metric\\
$\omega_{ab}$,$\epsilon_{ab}$ &-
&Metric/Levi-Civita pseudo-tensor on unit sphere\\
\hline

vertical bar ( $|$ )&- 
&Covariant derivative associated with $\bar\gamma_{ij}$\\
colon ( $:$ )&\eqref{eq:intrinsic covariant derivative} 
&Covariant derivative associated with $\omega_{ab}$\\
\hline

$\chi$&-
&Affine parameter on unperturbed spacetime\\
$\dd x^\mu /\dd\chi$&\eqref{eq:background wave vector}
&Wave vector on unperturbed spacetime \\
$\hat{\bm n}$&-&
line-of-sight vector\\
${\bm e}_a,{\bm e}_\pm$&\eqref{eq:polarization basis}
&Basis of spin-weight $\pm 1$\\
\hline

$\Phi ,\Psi$&\eqref{eq:Phi Psi def} 
&Gauge-invariant scalar metric perturbations\\
$\sigma_{\rmg ,i}$&\eqref{eq:sigma_g,i def}
&Gauge-invariant vector metric perturbations\\
$h_{ij}$&\eqref{eq:delta g_ij def}
&Tensor metric perturbations\\
$\Upsilon$&\eqref{eq:Upsilon def}
&Spin-$0$ gauge-invariant combination\\
$\Omega_a$&\eqref{eq:Omega_a def}
&Spin-$\pm 1$ gauge-invariant combination\\
$\Psi_{\bm k}^{(m)}$ &\eqref{eq:Psi_k^m}&
Fourier coefficients for mode-$m$ gauge-invariant perturbations\\ 
\hline

$\Delta^a$&\eqref{eq:deflection angle}
&Deflection angle on unit sphere \\
$\phi ,\varpi$&\eqref{eq:gradient-curl decomposition} 
&Scalar/pseudo-scalar lensing potentials\\
$\mcD^a{}_b$&\eqref{eq:Jacobi map def} 
&Jacobi map \\
$\mcT^a{}_b$&\eqref{eq:Jacobi map eq} 
&Symmetric optical tidal matrix \\
$\gamma_{ab}$&\eqref{eq:linear-order Jacobi map}
&shear field\\
$E,B$&\eqref{eq:reduced shear Y_lm expansion}
&E-/B-mode reduced cosmic shear\\
\hline

${}_sG_\ell{}^m$&\eqref{eq:TAM wave}
&basis function\\
${}_s\epsilon_L^{(\ell ,m)},{}_s\beta_L^{(\ell ,m)}$&\eqref{eq:epsilon beta def}
&Radial E,B function\\
$Q^{(0)},Q^{(\pm 1)}_i,Q^{(\pm 2)}_{ij}$&\eqref{eq:Q^(0) def},\eqref{eq:Q^(pm 1)_i def},\eqref{eq:Q^(pm 2)_ij def}
&spin-$0$,$\pm 1$,$\pm 2$ basis\\
$\pspin ,\mspin$&\eqref{eq:pspin def},\eqref{eq:mspin def}
&spin-raising/lowering operators\\
\hline

$\mcS_{x,\ell}^{(m)}$ &\eqref{eq:S_phi,ell^(0)}-\eqref{eq:S_varpi,ell^(2)}
& Transfer function for gradient-/curl-modes \\
$\mcS_{\rmX ,\ell}^{(m)}$ &\eqref{eq:S_E,ell^(0)}-\eqref{eq:S_B,ell^(2)}
& Transfer function for E-/B-modes \\
$\mcM_\ell^{\rmX\rmX}{}^{(m)}$ &\eqref{eq:M^xx(m) def},\eqref{eq:M^XX(m) def}
& Auto-power spectrum for $X_\ell{}^{(m)}({\bm k})$ \\
$C_\ell^{\rmX\rmX}$&\eqref{eq:C_l^xx' def},\eqref{eq:C_l^XX' def}
& Angular power spectrum for $X_{\ell m}$\\
$\mcP_{|m|}$&\eqref{eq:P_PsiPsi def}
& Auto-power spectrum for $\Psi_{\bm k}^{(m)}$\\
\hline 
\end{tabular}
\label{notation}
\end{table} 

\section{Background and perturbations}
\label{sec:Background and perturbations}

In this paper, we consider the flat FLRW universe 
with the metric given by
\al{
	\dd s^2
		=a^2(\eta )\tilde g_{\mu\nu}\dd x^\mu\dd x^\nu
		=&a^2(\eta )
			\bigl( g_{\mu\nu}+\delta g_{\mu\nu}\bigr)\dd x^\mu\dd x^\nu
	\,,\label{eq:metric def}
}
where $a(\eta )$ corresponds to the conventional scale factor of
a homogeneous and isotropic universe, $\tilde g_{\mu\nu}$ is the conformal flat
four-dimensional metric which includes the spacetime inhomogeneity, 
$\delta g_{\mu\nu}$ is small metric perturbations.
Here, $g_{\mu\nu}$ is the conformally related metric assumed to have 
the following form: 
\al{
	g_{\mu\nu}\dd x^\mu\dd x^\nu
		=-\dd\eta^2 +\bar\gamma_{ij}\dd x^i\dd x^j
		=-\dd\eta^2 +\dd\chi^2 +\chi^2\omega_{ab}\dd\theta^a\dd\theta^b
	\,,\label{eq:unperturbed 4-dimensional metric}
}
where $\omega_{ab}=\dd\theta^2 +\sin^2\theta\dd\varphi^2$ is the metric
on the unit sphere.
In what follows, tensors defined
on the perturbed spacetime will be distinguished by
the indication of a tilde ( $\tilde{}$ ) as above.
The small departure of the metric from the background metric $g_{\mu\nu}$
can be represented as a set of metric perturbations:
\al{
	&\delta g_{00}=-2A
	\,,\\
	&\delta g_{0i}=B_{|i}+B_i
	\,,\\
	&\delta g_{ij}=2\mcR\bar\gamma_{ij}+2H_{|ij}+H_{i|j}+H_{j|i}+h_{ij}
	\,,\label{eq:delta g_ij def}
}
where $B_i$ and $H_i$ are divergence-free three-vectors,
$h_{ij}$ is the transverse-traceless tensor, and the vertical bar ( $|$ )
denotes the covariant derivative with respect to the three-dimensional metric
$\bar\gamma_{ij}$\,.

Based on the gauge transformation properties, independent gauge-invariant quantities
can be constructed from these variables.
One possible choice of such invariants are~\cite{Kodama:1985bj}
\al{
	&\Phi\equiv A-\frac{1}{a}\left( a\left(\dot H-B\right)\right)^{\cdot}
	\,,\ \ \ 
	\Psi\equiv \mcR -\frac{\dot a}{a}\left(\dot H-B\right)
	\,,\label{eq:Phi Psi def}
}
where the dot ( $\dot{}$ ) denotes the derivative with respect to the conformal
time $\eta$\,.
These combinations corresponds to the Bardeen's invariants
and they are chosen as the appropriate variables for the conformal Newton-gauge.
We also have the gauge-invariant vector metric perturbations: \cite{Kodama:1985bj}
\al{
	&\sigma_{\rmg ,i}\equiv\dot H_i-B_i
	\,.\label{eq:sigma_g,i def}
}
For tensor perturbations, there exists no tensor-type infinitesimal gauge
transformation.
Hence all the quantities associated with tensor perturbations are
gauge-invariant by themselves.
Appendix \ref{sec:Christoffel symbols and Riemann tensors} summarizes
the Christoffel symbols and Riemann tensors from the metric perturbations $\delta g_{\mu\nu}$\,. 

\bc
\begin{figure}[tbp]
\bc
\includegraphics[width=100mm]{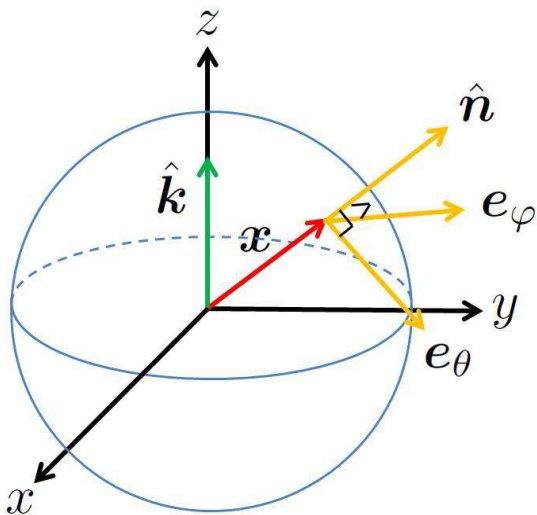}
\caption{
Representation of the direction vector $\hat{\bm n}$\,, and 
the two orthogonal vectors along the light ray $\{{\bm e}_\theta\,,{\bm e}_\phi\}$\,.
At the observer position, these vectors form a basis, which can be parallel transported
along the geodesic.
}
\label{fig:coordinate_system}
\ec
\end{figure} 
\ec

We consider two null geodesics on the physical spacetime : $x^\mu (v)$ 
and $\tilde x^\mu (v)=x^\mu (v)+\xi^\mu (v)$\,, where $v$ is the affine parameter 
along the photon path and $x^\mu (v)$ and $\xi^\mu (v)$ are 
a reference geodesic and a deviation vector labeling the reference geodesic.
It is known that the conformal transformation $a^2g_{\mu\nu}\rightarrow g_{\mu\nu}$
maps a null geodesic on the physical spacetime to a null geodesic on
the conformally transformed spacetime with the affine parameter
transformed as $\dd v\rightarrow \dd\lambda =a^{-2}\dd v$ \cite{Wald,Sasaki:1987ad}\,.
In the cosmological background, it is sufficient to perform
the calculation without the Hubble expansion and reintroduce
the scale factor at the end by scaling $\dd\lambda\rightarrow\dd v=a^2\dd\lambda$\,.
Hence, we define a tangent vector $k^\mu$ on the conformally transformed spacetime as
\al{
	k^\mu \equiv a^2\frac{\dd x^\mu}{\dd v} =\frac{\dd x^\mu}{\dd\lambda}
	\,.\label{eq:k def}
}
This is a null vector satisfying the equations:
\al{
	g_{\mu\nu}k^\mu k^\nu =0
	\,,\ \ \ 
	\frac{\DD k^\mu}{\dd\lambda}
		\equiv\frac{\dd^2 x^\mu}{\dd\lambda^2}
			+\Gamma^\mu_{\rho\sigma}
			\frac{\dd x^\rho}{\dd\lambda}\frac{\dd x^\sigma}{\dd\lambda}
		=0
	\,,\label{eq:tilde k^mu equations}
}
where $\Gamma^\mu_{\rho\sigma}$ is the Christoffel symbols associated with
the metric on the unperturbed universe, $g_{\mu\nu}$\,.
We can solve the above geodesic equation to obtain
$x^\mu (\lambda )=E(\lambda ,(\lambda_0 -\lambda )\hat{\bm n})$\,,
where $E$ and $\hat{\bm n}$ represent the photon energy and the propagation
direction measured from the observer in the background flat spacetime\,,
$\lambda_0$ denotes the affine parameter at the observer.
$\hat{\bm n}$ is the unit vector tangent to geodesic on the flat three-space,
satisfying $\hat{\bm n}\cdot\hat{\bm n}=1$ and $\hat n^i{}_{|j}\hat n^j=0$\,.
Here we have switched from $\lambda$ to $\chi\equiv E (\lambda_0 -\lambda )$\,.
We then have the wave vector in the unperturbed universe:
\al{
	&\frac{\dd x^\mu}{\dd\chi}
		=\left( -1\,,\,\hat{\bm n}\right)
	\,.\label{eq:background wave vector}
}
We also denote by $u^\mu$ the background observer's four-velocity
at the observer's position, with the normalization condition 
$g_{\mu\nu}u^\mu u^\nu =-1$\,.
With these notations, we define orthogonal spacetime basis along the light ray, 
$e^\mu_a$ with $a=\theta\,,\varphi$\,, which obey
\al{
	g_{\mu\nu}e^\mu_a e^\nu_b =\omega_{ab}
	\,,\ \ 
	g_{\mu\nu}k^\mu e^\nu_a 
		=g_{\mu\nu} u^\mu e^\nu_a =0
	\,.
}
They are parallel transported along the geodesics : $(\DD /\dd\chi )u^\mu =0$\,,
$(\DD /\dd\chi )e^\mu_a=0$\,.
Representation of light bundle, the deviation vector,
and the basis vectors is shown in fig.~\ref{fig:coordinate_system}.
Considering a static observer, $u^\mu =(1,{\bm 0})$\,, the basis vector
$e^\mu_a$ can be described as $e^\mu_a =(0,{\bm e}_a)$\,.
To discuss the spatial pattern on celestial sphere, it is useful to introduce
the spin-weighted quantities.
The polarization vector with respect to a two-dimensional vector on the sky
is expected in terms of two vector basis ${\bm e}_\theta (\hat{\bm n})$ and ${\bm e}_\varphi (\hat{\bm n})$
perpendicular to the line-of-sight vector ${\bm n}$ as \cite{Lewis:2001hp}
\al{
	{\bm e}_\pm (\hat{\bm n})
		={\bm e}_\theta (\hat{\bm n})\pm\frac{i}{\sin\theta}\,{\bm e}_\varphi (\hat{\bm n})
	\,.\label{eq:polarization basis}
}
We list the explicit expression for the basis vectors, $\hat{\bm n}$ and ${\bm e_a}$\,, 
in the Cartesian coordinates in Appendix \ref{sec:Intrinsic covariant derivative}.

\section{Weak lensing observables}
\label{sec:Weak lensing observables}

In this section, we consider the deformation of the cross-section 
of a congruence of null geodesics under propagation in a perturbed universe.
We give the basic equations which govern the weak 
gravitational lensing effect in the presence of scalar, vector and 
tensor metric perturbations 
by solving the geodesic equation in section \ref{sec:Deflection angle} 
and the geodesic deviation equation in section \ref{sec:Jacobi map}.
Based on the results, we derive an explicit relation between 
the gradient of the deflection angle and the Jacobi matrix.

\subsection{Deflection angle}
\label{sec:Deflection angle}
 
In order to see the lensing effect, let us consider the spatial components
of the geodesic equation for the photon ray.
To derive the first-order geodesic equation, we parametrize the perturbed photon geodesic as
\al{
	\frac{\dd\tilde x^\mu}{\dd\chi}
		=\left( -1+\delta\nu\,,\,\hat{\bm n}+\delta{\bm e}\right)
	\,.
	\label{eq:perturbed wave vector}
}
Based on the gauge transformation properties, we can construct 
the gauge-invariant components of perturbed wave vector~\cite{Yoo:2010ni}:
\al{
	&\delta\nu_{\rm GI} 
		=\delta\nu 
			+2\frac{\dot a}{a}\sigma_\rmg^{(\rmS )}
			-\frac{\dd}{\dd\chi}\sigma_\rmg^{(\rmS )}
	\,,\label{eq:gauge inv dn}\\
	&\delta e_{\rm GI}^i 
		=\delta e^i +2\frac{\dot a}{a}\sigma_\rmg^{(\rmS )}\hat n^i
			+\frac{\DD}{\dd\chi}\mcG^i -\mcG^j\hat n^i{}_{|j}
	\,,\label{eq:gauge inv de}
}
where $\sigma_\rmg^{(\rmS )}\equiv \dot H-B$ and $\mcG_i\equiv (H_{|i}+H_i)$ 
are the pure gauge terms.
They are chosen as the appropriate variables for the conformal Newton gauge.
To derive the first-order geodesic equation, we expand the Christoffel symbols as
$\tilde\Gamma^\mu_{\rho\sigma}=\Gamma^\mu_{\rho\sigma}+\delta\Gamma^\mu_{\rho\sigma}$\,,
where $\tilde\Gamma^\mu_{\rho\sigma}$ is the Christoffel symbols associated
with the metric on the perturbed universe, 
$\tilde g_{\mu\nu}$ (see Appendix \ref{sec:Christoffel symbols and Riemann tensors})\,.
In the conformally transformed spacetime, 
we then derive the null condition, the temporal and spatial components of the geodesic equation 
in terms of the gauge-invariant quantities:
\al{
	&\hat{\bm n}\cdot\delta{\bm e}_{\rm GI}
		=\delta\nu_{\rm GI}-\Upsilon
	\,,\ \ \ \ 
	\frac{\dd}{\dd\chi}
		\Bigl(
			\delta\nu_{\rm GI} +2\Phi -\sigma_{\rmg ,i}\hat n^i
		\Bigr)
		=\dot\Upsilon
	\,,\\
	&\frac{\DD}{\dd\chi}
		\Bigl(
			\delta e^i_{\rm GI}
			+2\Psi\hat n^i +\sigma_\rmg{}^i+h^i{}_j\hat n^j
		\Bigr)
		=-\delta e^j_{\rm GI}\,\hat n^i{}_{|j}+\left(\Psi -\Phi\right)^{|i}
			+\sigma_{\rmg\, j}{}^{|i}\hat n^j 
			+\frac{1}{2}h_{jk}{}^{|i}\hat n^j\hat n^k
	\,,\label{eq:perturbed spacelike geodesic eq}
}
where $(\DD /\dd\chi )\zeta^i\equiv\hat n^j\zeta^i{}_{|j} -\pd_\eta\zeta^i$ and
we have introduced the gauge-invariant combination of the spin-$0$ components
constructed from the scalar, vector, and tensor perturbations as
\beq
	\Upsilon
		\equiv\Psi -\Phi +\sigma_{\rmg ,i}\,\hat n^i+\frac{1}{2}\,h_{ij}\,\hat n^i\hat n^j
	\,.\label{eq:Upsilon def}
\eeq
The equations we have derived here are manifestly gauge-invariant 
because the gauge degrees of freedoms in the explicit expressions 
for the geodesic equation are completely canceled.

In addition to the wave vector, we introduce the gauge-invariant deviation
vector as $\delta e^i_{\rm GI}\equiv (\dd /\dd\chi )\xi^i_{\rm GI}$\,.
To extract the angular components of the deviation vector, $\xi^a_{\rm GI}\equiv\xi^i_{\rm GI}e_i^a$, 
we multiply  $e_i^a$ in both side of eq.~\eqref{eq:perturbed spacelike geodesic eq}\,.
Since the unperturbed Christoffel symbols satisfies $\Gamma^i_{jk}=0$
in the Cartesian coordinate system, with the condition for the parallel transformation, 
$(\DD /\dd\chi )e_i^a=0$\,, 
we obtain
\al{
	&\frac{\dd^2\xi^a_{\rm GI}}{\dd\chi^2}
		=\frac{1}{\chi}\omega^{ab}
			\biggl\{
			 \Upsilon_{:b}
			 -\frac{\dd}{\dd\chi}
				\bigl(
					\chi\Omega_b
				\bigr)
			\biggr\}
	\,.
}
where we have introduced the gauge-invariant combination of the spin-$\pm 1$ components 
constructed from the vector and tensor metric perturbations:
\al{
	\Omega_a\equiv\left(\sigma_{\rmg ,j}+h_{ij}\hat n^i\right) e^j_a
	\,.\label{eq:Omega_a def}
}
Given the initial conditions, $\xi^a_{\rm GI}|_0 =0$\,, 
and $(\dd\xi^a_{\rm GI} /\dd\chi )|_0 =\delta\theta_0^a$\,, where
$\delta\theta^a_0$ denotes the angular coordinate at the observer position, 
the deviation vector as
\al{
	\frac{\xi^a_{\rm GI}}{\cS}
		=&\,\delta\theta^a_0 
			+\omega^{ab}
				\int^{\cS}_0\dd\chi\frac{\cS -\chi}{\cS\chi}
				\biggl\{
					\Upsilon_{:b}
					-\frac{\dd}{\dd\chi}
						\bigl(
							\chi\Omega_b
						\bigr)
				\biggr\}\biggl|_{(\eta_0 -\chi ,\chi\hat{\bm n})}
	\,.\label{eq:deviation vector}
}
The integration at the right-hand-side is evaluated along the unperturbed light path
$x^\mu (\chi )=(\eta_0 -\chi ,\chi\hat{\bm n})$\,, 
where $\eta_0$ denotes the conformal time at the observer, 
according to the Born approximation.
For simplicity, we omit the subscript $(\eta_0 -\chi ,\chi\hat{\bm n})$ hereafter.
Given the angular direction at both end points, the deflection angle, $\Delta^a$, 
can be estimated through \cite{Stebbins:1996wx}
\al{
	\Delta^a\equiv\frac{\xi^a_{\rm GI}}{\cS} -\delta\theta^a_0
	\,.\label{eq:deflection angle}
}
Since the deflection angle is the two-dimensional vector field defined on 
the celestial sphere, it is generally characterized by the sum of
two potentials;
a gradient of scalar lensing potential ($\phi$) (gradient-mode),
and a rotation of pseudo-scalar lensing potential ($\varpi$)
(curl-mode)~\cite{Hirata:2003ka}:
\al{
	\Delta^a =\phi_{:a}+\varpi_{:b}\epsilon^b{}_a
	\,.\label{eq:gradient-curl decomposition} 
}
where $\epsilon^b{}_a$ denotes the two-dimensional Levi-Civita pseudo-tensor.
Integrating by part, the gradient-/curl-mode lensing potentials 
can be written as
\al{
	&\phi^{:a}{}_{:a} 
		=\Delta^a{}_{:a}
		=\int^{\cS}_0\dd\chi\frac{\cS -\chi}{\cS\chi}
				\biggl\{
					\Upsilon^{:a}{}_{:a}
					-\frac{\dd}{\dd\chi}\left(\chi\Omega^a{}_{:a}\right)
				\biggr\}
	\,,\label{eq:scalar lensing potential}\\
	&\varpi^{:a}{}_{:a}
		=\Delta^a{}_{:b}\epsilon^b{}_a
		=-\int^{\cS}_0\dd\chi\frac{\cS -\chi}{\cS\chi}
			\frac{\dd}{\dd\chi}\left(\chi\Omega^a{}_{:b}\epsilon^b{}_a\right)
	\,.\label{eq:pseudo-scalar lensing potential}
}

\subsection{Jacobi map}
\label{sec:Jacobi map}

Let us consider the Jacobi map which characterizes the deformation of light bundle.
In terms of the projected deviation vector $\xi^a$\,,
the geodesic deviation equation in the conformally transformed spacetime 
can be written as~\cite{Seitz:1994xf,1961RSPSA.264..309S}
\al{
	\frac{\dd^2\xi^a}{\dd\chi^2}
		=\tilde\mcT^a{}_b\,\xi^b
	\,;\ \ 
	\tilde\mcT^a{}_b=-\tilde R_{\mu\rho\nu\sigma}
					\frac{\dd x^\mu}{\dd\chi}\frac{\dd x^\nu}{\dd\chi}
					e^{\rho a}e^\sigma_b
	\,,\label{eq:geodesic deviation equation}
}
where $\tilde\mcT^a{}_b$ is the symmetric optical tidal matrix, 
and $\tilde R_{\mu\rho\nu\sigma}$
is the Riemann tensor associated with the metric $\tilde g_{\mu\nu}$\,.
Provided the initial conditions at the observer, $\xi^a|_0 =0$
and $(\dd\xi^a /\dd\chi )|_0 =\delta\theta_0^a$\,,
the solution of eq.~\eqref{eq:geodesic deviation equation} can be
written in terms of the Jacobi map as
\al{
	\xi^a =\tilde\mcD^a{}_b\,\delta\theta^b_0
	\,,\label{eq:Jacobi map def} 
}
where the Jacobi map $\tilde\mcD^a{}_b$ satisfies
\al{
	\frac{\dd^2}{\dd\chi^2}\tilde\mcD^a{}_b =\tilde\mcT^a{}_c\tilde\mcD^c{}_b
	\,,\label{eq:Jacobi map eq}
}
with the initial condition for the Jacobi map : $\tilde\mcD^a{}_b|_0 =0$
and $(\dd /\dd\chi )\tilde\mcD^a{}_b|_0 =\delta^a{}_b$\,.

To obtain the expression relevant for the weak lensing measurements,
we expand as $\tilde\mcD^a{}_b =\mcD^a{}_b +\delta\mcD^a{}_b$ and 
$\tilde\mcT^a{}_b =\mcT^a{}_b +\delta\mcT^a{}_b$\,.
Since $\mcT^a{}_b=0$ in unperturbed spacetime,
the zeroth-order solution of Jacobi map trivially reduces to
$\mcD^a{}_b=\chi\delta^a{}_b$\,.
Substituting this expression into eq.~\eqref{eq:Jacobi map eq}
and solving this equation, we have the Jacobi map
up to linear order:
\al{
	\frac{1}{\cS}\tilde\mcD^a{}_b
		=\delta^a{}_b
			+\int^{\cS}_0\dd\chi\frac{\left(\cS -\chi\right)\chi}{\cS}\,
			\delta\mcT^a{}_b (\eta_0 -\chi ,\chi\hat{\bm n})
			+\mcO (\delta g_{\mu\nu}^2)
	\,.\label{eq:first order Jacobi map sol}
}

We are now interested in the shear fields, namely the symmetric trace-free part
of the Jacobi map.
We introduce the bracket $\ave{\cdots}$\,, which denotes 
the symmetric trace-free part taken in the two-dimensional space:
$X_{\ave{ab}}=(1/2)(X_{ab}+X_{ba}-X^c{}_c\,\omega_{ab})$\,.
To derive the expression relevant for the arbitrary metric perturbations 
eq.~\eqref{eq:metric def}, we explicitly write down
the symmetric trace-free part of the linear-order symmetric optical tidal 
matrix as (see Appendix \ref{sec:Christoffel symbols and Riemann tensors})
\al{	
	\chi^2\,\delta\mcT_{\ave{ab}}
		=\Upsilon_{\ave{:ab}}
			-\frac{\dd}{\dd\chi}
				\Bigl(
					\chi\Omega_{\ave{a:b}}
				\Bigr)
			+\frac{1}{2}\chi\frac{\dd^2}{\dd\chi^2}\left(\chi h_{\ave{ab}}\right)
	\,,\label{eq:deltaT}
}
where $\Upsilon$ and $\Omega_a$ were defined in 
eqs.~\eqref{eq:Upsilon def} and \eqref{eq:Omega_a def}\,,
we have defined $h_{ab}=h_{ij}e^i_a e^j_b$\,.
Substituting eq.~\eqref{eq:deltaT}
into eq.~\eqref{eq:first order Jacobi map sol} and integrating by part,
we obtain the explicit expression for the first-order cosmic shear as
\al{
	\gamma_{ab}
		\equiv &\frac{1}{\cS}\tilde\mcD_{\ave{ab}}
		=\int^{\cS}_0\dd\chi\frac{\cS -\chi}{\cS\chi}
			\biggl\{
				\Upsilon_{:\ave{ab}}
					-\frac{\dd}{\dd\chi}\left(\chi\Omega_{\ave{a:b}}\right)
			\biggr\}
			+\frac{1}{2}\bigl[h_{\ave{ab}}\bigr]^{\cS}_0
	\,,\label{eq:linear-order Jacobi map}
}
where $[\zeta ]^{\chi_1}_{\chi_2}\equiv\zeta (\eta_0 -\chi_1 ,\chi_1\hat{\bm n})-\zeta (\eta_0 -\chi_2 ,\chi_2\hat{\bm n})$\,.
Since the gauge degrees of freedom are completely removed in the explicit expression
for the symmetric optical tidal matrix \eqref{eq:deltaT},
the resulting shear field we have derived here are manifestly gauge-invariant.
\footnote{
We should comment on the effect of the velocity. 
Although the Jacobi map is in general expected to depend on the velocity at the source
and observer, these contributions appears only in the trace-part of the Jacobi map, namely
convergence field (see \cite{Schmidt:2012ne}).
Hence the shear we have derived here does not include the effect of the velocity.}

Finally, comparing between \eqref{eq:deviation vector}, 
\eqref{eq:gradient-curl decomposition}, and \eqref{eq:linear-order Jacobi map}\,,
we find the explicit relation between the deflection angle
and the deformation matrix:
\al{
	\gamma_{ab}
		=&\Delta_{\ave{a:b}}
			+\frac{1}{2}
				\bigl[h_{\ave{ab}}\bigr]^{\cS}_0
	\,.\label{eq:gamma Delta relation}
}
This relation is one of the main results in this paper.
Although cosmic shear measurement via galaxy survey are usually 
referenced to the coordinate in which galaxies are statistically 
isotropic, this is in general different from our reference coordinate, 
namely flat FLRW universe.
Hence, the correction from the gravitational potential at the source
should appear in the observed shear field.
Such correction corresponds to the last term at the right-hand-side 
in eq.~\eqref{eq:gamma Delta relation} and
is referred to as the metric shear/Fermi Normal Coordinate (FNC) term, 
which has been discussed
in Refs.~\cite{Dodelson:2003bv,Schmidt:2012ne,Schmidt:2012nw}.
In contrast to the previous studies based on the geodesic equation, 
the metric shear/FNC term naturally arises in our case 
from the geodesic deviation equation. 
This is understood as follow:
Since the leading correction of the metric in the FNC is known to 
be described by the Riemann curvature, it contains the information
of the difference between the FNC and the flat FLRW coordinates.
Therefore the FNC contribution is automatically included in the symmetric tidal 
matrix perturbed around the flat FLRW universe.
Furthermore the geodesic equation contains only up to
the first line-of-sight derivative on the metric perturbations, 
while the metric shear/FNC term is a second-rank tensor, which 
appears only through the second line-of-sight derivative on the metric
perturbations (see eq.~\eqref{eq:deltaT}). 

\section{Total angular momentum method}
\label{sec:Total angular momentum method}

In this section, we introduce the TAM representation
for the fluctuation modes to derive the full-sky formulae of angular power 
spectra for the reduced shear and the deflection angle.  
Hereafter we follow and extend the formalism developed by \cite{Hu:1997hp}
(see also \cite{Dai:2012bc}).
Using the gauge degrees of freedom for scalar and vector perturbations,
we adopt the conformal Newton gauge: $B=H=0$ and $H_i=0$\,.
In section \ref{sec:Basis function}\,, we first introduce
the basis function for spin-$s$ field with a given Fourier mode 
and its dependence is summarized.
We review the decomposition of metric perturbations into scalar,
vector, and tensor modes and present the relationship between the mode function
and the basis function in section \ref{sec:Mode functions}.
In sections \ref{sec:Gradient- and curl-modes} and \ref{sec:E- and B-modes} 
we present the formula for the angular power spectrum
for the gradient-/curl-mode lensing potential and the E-/B-mode cosmic shear
generated by scalar, vector, and tensor perturbations.

\subsection{Basis function}
\label{sec:Basis function}

Weak lensing observables are in general
functions of both spatial position ${\bm x}$ and angle $\hat{\bm n}$\,.
The fluctuations can be decomposed into the harmonic modes which are 
the eigenfunction for the Laplace operator.
For a fixed Fourier mode ${\bm k}$\,,  the plane wave 
$e^{-i{\bm k}\cdot{\bm x}}$ form a complete basis 
in the three-dimensional flat space.
The spin-$s$ field for a given Fourier mode ${\bm k}$ 
generally may be expanded in
\al{
	{}_sG_\ell{}^m({\bm x},\hat{\bm n},{\bm k})
		\equiv &(-i)^\ell\sqrt{\frac{4\pi}{2\ell +1}}\,
				{}_sY_\ell{}^m(\hat{\bm n})\, e^{-i{\bm k}\cdot{\bm x}}
	\,.\label{eq:TAM wave}
}
Without loss of generality, we can choose coordinate system with 
${\bm k}\parallel {\bm e}_z$\,.
In this coordinate system, the orbital angular momentum of the plane wave can be written as
a sum of $L$:
\al{
	e^{-i{\bm k}\cdot{\bm x}}
		=\sum_{L=0}^\infty (-i)^L\sqrt{4\pi (2L+1)}j_L(k\chi )Y_L{}^0(\hat{\bm n})
	\,.
}
where we have used ${\bm x}=\chi\hat{\bm n}$\,. 
Hence, the basis function ${}_sG_\ell{}^m ({\bm x},\hat{\bm n},{\bm k})$ for fixed ${\bm k}$
can be decomposed into their total angular momentum components:
\al{
	{}_sG_\ell{}^m({\bm x},\hat{\bm n},{\bm k})
		=&\sum_{L=0}^\infty 4\pi\sqrt{\frac{2L+1}{2\ell +1}}i^{-L-\ell}
			j_L(k\chi )Y_L{}^0(\hat{\bm n})\,{}_sY_\ell{}^m(\hat{\bm n})
	\notag\\
		=&\sum_{L=0}^\infty 
			(-i)^L\sqrt{4\pi (2L+1)}\,
			\left({}_s\epsilon_L^{(\ell ,m)}(k\chi )+i\,{\rm sgn}(s)\,{}_s\beta_L^{(\ell ,m)}(k\chi )\right)\,
			{}_sY_L{}^m(\hat{\bm n})
	\,,\label{eq:TAM wave expand}
}
where we have used the Clebsch-Gordan relation (see eq.~\eqref{eq:Clebsch-Gordan relation}).
Here ${\rm sgn}(s)$ denotes the signature of $s$.
We define the functions ${}_s\epsilon_L^{(\ell ,m)}$ and ${}_s\beta_L^{(\ell ,m)}$
which represent the sums over $j$ \cite{Durrer}
\al{
	{}_s\epsilon_L^{(\ell ,m)}(x)+i\,{\rm sgn}(s)\,{}_s\beta_L^{(\ell ,m)}(x)
		=\sum_{j=|L-\ell |}^{L+\ell}(-i)^{j+\ell -L}\frac{2j+1}{2L+1}
			\langle\, j,\ell ;0,m|L,m\,\rangle\langle\, j,\ell ;0 ,-s|L,-s\,\rangle j_j(x)
	\,,\label{eq:epsilon beta def}
}
where $\langle\ell_1 ,\ell_2;m_1,m_2|\ell ,m\rangle$ denotes
the Clebsch-Gordan coefficient.
Appendix \ref{sec:Eplicit expression} summarize 
the explicit expression for ${}_s\epsilon_L^{(\ell ,m)}$
and ${}_s\beta_L^{(\ell ,m)}$ with $s=0,\pm 1,\pm2$\,.

\subsection{Mode functions}
\label{sec:Mode functions}

In this subsection, we briefly review the properties of
the mode functions for scalar, vector, and tensor perturbations.
We then see that scalars, vectors, and tensors generates
only $m=0\,,1$\,, and $2$ fluctuations, respectively.

\subsubsection{Scalar mode} 

The scalar mode function, $Q^{(0)}$\,, is the eigenfunction for the Laplace operator 
on the three-dimensional flat space:
\al{
	Q^{(0)}{}^{|i}{}_{|i}=-k^2Q^{(0)}\,.
}
This is represented by the plane wave:
\al{
	Q^{(0)}
		=e^{-i{\bm k}\cdot{\bm x}}
	\,.\label{eq:Q^(0) def}
}
Notice that in the coordinate system with ${\bm k}\parallel{\bm e}_z$\,, 
one can easily show that the mode function can be described by
the basis function as $Q^{(0)}={}_0G_0{}^0$\,.
With a help of the scalar harmonics, the scalar metric perturbations 
$\Phi$ and $\Psi$ are expanded as
\al{
	&\Phi ({\bm x},\eta )
		=\int\frac{\dd^3{\bm k}}{(2\pi )^3}\Phi_{\bm k} (\eta )\,Q^{(0)}({\bm x},{\bm k})
	\,,\ \ \ 
	\Psi ({\bm x},\eta )
		=\int\frac{\dd^3{\bm k}}{(2\pi )^3}\Psi_{\bm k} (\eta )\,Q^{(0)}({\bm x},{\bm k})
	\,.
}

\subsubsection{Vector mode} 

Vector perturbations can be decomposed into the mode function, $Q^{(\pm 1)}_i$\,,
which is the eigenfunction of the Laplace operator in the same manner
as the scalar mode:
\al{
	&Q_i^{(\pm 1)}{}^{|j}{}_{|j}=-k^2Q_i^{(\pm 1)}
	\,,\ \ \ 
	Q_i^{(\pm 1)}{}^{|i}=0
	\,.
}
In the coordinate system with ${\bm k}\parallel{\bm e}_z$\,, 
a convenient representation 
for the vector mode with a given Fourier mode ${\bm k}$ would be
\al{
	&Q_i^{(\pm 1)}=\pm\frac{i}{\sqrt{2}}e_{\pm ,i}(\hat{\bm k})\, e^{-i{\bm k}\cdot{\bm x}}
	\,,\label{eq:Q^(pm 1)_i def}
}
where ${\bm e}_\pm (\hat{\bm k})$ denotes the polarization vector perpendicular
to $\hat{\bm k}$ (see eq.~\eqref{eq:polarization basis})\,.
Using the properties of the spin-weighted spherical harmonics 
(see Appendix \ref{sec:Spherical harmonics}), 
we have the relation between the mode function and the basis function as
\al{
	\hat{\bm n}\cdot{\bm Q}^{(m)}={}_0G_1{}^{m}
	\,,\ \ \ 
	{\bm e}_\pm (\hat{\bm n})\cdot{\bm Q}^{(m)}
		=\mp\sqrt{2}\,{}_{\pm 1}G_1{}^{m}
	\,,
}
for $m=\pm 1$\,.
Using the vector harmonics, the vector metric perturbations are expanded as
\al{
	\sigma_{\rmg ,i}({\bm x},\eta )
		=\int\frac{\dd^3{\bm k}}{(2\pi )^3}
			\sum_{m=\pm 1}\sigma_{\rmg ,{\bm k}}^{(m)}(\eta )\,Q_i^{(m)}({\bm x},{\bm k})
	\,.
}

\subsubsection{Tensor mode} 

In the same manner as the scalar and vector modes,
tensor mode functions are represented by Laplace eigenfunctions:
\al{
	&Q_{ij}^{(\pm 2)}{}^{|k}{}_{|k}=-k^2Q_{ij}^{(\pm 2)}
	\,,\ \ \ 
	{\rm Tr}\left( Q_{ij}^{(\pm 2)}\right) =0
	\,,\ \ \ 
	Q_{ij}^{(\pm 2)}{}^{|i}=0
	\,.
}
With a help of the spin-weight $\pm 1$ polarization vector, ${\bm e}_\pm (\hat{\bm k})$\,,
we obtain the explicit expression as:
\al{
	&Q_{ij}^{(\pm 2)}
		=-\frac{1}{\sqrt{2}}\,
			e_{\pm ,i}(\hat{\bm k})e_{\pm ,j}(\hat{\bm k})\,
			e^{-i{\bm k}\cdot{\bm x}}
	\,.\label{eq:Q^(pm 2)_ij def}
}
With these notations and the properties of the spin-weighted spherical harmonics 
(see Appendix \ref{sec:Spherical harmonics}), 
in the coordinate system with ${\bm k}\parallel{\bm e}_z$\,, 
one can verify
\al{
	&\hat n^i\hat n^jQ_{ij}^{(m)}
		=\frac{2}{\sqrt{3}}\,{}_0G_2{}^m
	\,,\ \ 
	\hat n^ie_\pm^j (\hat{\bm n})\,Q_{ij}^{(m)}
		=\mp\sqrt{2}\,{}_{\pm 1}G_2{}^m
	\,,\\
	&e_\pm^i (\hat{\bm n})e_\pm^j (\hat{\bm n})\,Q_{ij}^{(m)}
		=2\sqrt{2}\,{}_{\pm 2}G_2{}^m
	\,,
}
for $m=\pm 2$\,.
One can expand the tensor metric perturbations in terms of the tensor harmonics:
\al{
	h_{ij}({\bm x},\eta )
		=\int\frac{\dd^3{\bm k}}{(2\pi )^3}
			\sum_{m=\pm 2}h_{\bm k}^{(m)}(\eta )Q_{ij}^{(m)}({\bm x},{\bm k})
	\,.
}

\subsection{Gradient- and curl-modes}
\label{sec:Gradient- and curl-modes}

Based on the expression eqs.~\eqref{eq:scalar lensing potential}, 
\eqref{eq:pseudo-scalar lensing potential} and the TAM representation
developed in section \ref{sec:Basis function}, 
we derive the angular power spectrum for the gradient-/curl-mode 
lensing potentials. Since these potentials 
transform as
spin-$0$ fields, they are decomposed on the basis of spherical harmonics
\al{
	x=\sum_{\ell =1}^\infty\sum_{m=-\ell}^\ell 
			x_{\ell m}\,Y_\ell{}^m
	\,,
}
where $x=\phi\,,\varpi$\,.
The auto- and cross-power spectra of these quantities are defined as
\al{
	C_\ell^{xx'}
		\equiv\frac{1}{2\ell +1}\sum_{m=-\ell}^\ell\,\ave{x_{\ell m}^*x'_{\ell m}}
	\,,\label{eq:C_l^xx' def}
}
where $x\,,x'=\phi\,,\varpi$ and the angle bracket $\ave{\cdots}$
denotes the ensemble average.

Notice that the metric on the unit sphere, $\omega^{ab}$\,,
and the Levi-Civita pseudo-tensor, $\epsilon^{ab}$\,, can be
rewritten in terms of the basis vectors $e^a_\pm$ with \cite{Lewis:2006fu}
\al{
	\omega^{ab}
		=\frac{1}{2}\Bigl( e^a_+e^b_-+e^a_-e^b_+\Bigr)
	\,,\ \ \ 
	\epsilon^{ab}
		=\frac{1}{2}i\,\Bigl( e^a_+e^b_--e^a_-e^b_+\Bigr)
	\,.
}
In terms of the spin-raising/lowering operators defined in 
eqs.~\eqref{eq:pspin def}, \eqref{eq:mspin def},
the gradient-/curl-mode lensing potentials are recast as
\al{
	&\pspin\mspin\,\phi
		=\int^{\cS}_0\dd\chi\frac{\cS -\chi}{\cS\chi}
				\Biggl[
					\pspin\mspin\,\Upsilon
					+\frac{1}{2}
						\frac{\dd}{\dd\chi}
						\biggl\{
						\chi
						\Bigl(
							\mspin\bigl( e_+^a\Omega_a\bigr) 
							+\pspin\bigl( e_-^a\Omega_a\bigr)
						\Bigr)
						\biggr\}
				\Biggr]
	\,,\label{eq:pspin mspin phi}\\
	&\pspin\mspin\,\varpi
		=\frac{1}{2}i
				\int^{\cS}_0\dd\chi\frac{\cS -\chi}{\cS\chi}
					\frac{\dd}{\dd\chi}
					\biggl\{
						\chi
						\Bigl(
							\mspin\bigl( e_+^a\Omega_a\bigr) 
							-\pspin\bigl( e_-^a\Omega_a\bigr)
						\Bigr)
					\biggr\}
	\,,\label{eq:pspin mspin varpi}
}
where we have used the relations between the intrinsic covariant derivative and spin-operators:
${}_0X_{:ab}e^a_+e^b_-=\pspin\mspin\left({}_0X\right)$\,,
$X_{a:b}e^a_+e^b_-=-\mspin\left( e^a_+X_a\right)$\,, 
and $X_{a:b}e^a_-e^b_+=-\pspin\left( e^a_-X_a\right)$\,.
Using the relation between the basis function (see section \ref{sec:Basis function}) 
and the mode function (see section \ref{sec:Mode functions})\,,
we decompose the gauge-invariant combinations, $\Upsilon$ and ${}_{\pm 1}\Omega$\,, 
into the Fourier coefficients of the gauge-invariant scalar/vector/tensor perturbations:
\al{
	&\Upsilon (\eta_0 -\chi ,\chi\hat{\bm n})
		=\int\frac{\dd^3{\bm k}}{(2\pi )^3}
			\sum_{m=-2}^2\Upsilon_{\bm k}^{(m)}(\eta_0 -\chi )\,
			{}_0G_{|m|}{}^m(\chi\hat{\bm n} ,\hat{\bm n},{\bm k})
	\,,\label{eq:Theta expand}\\
	&e^a_\pm\Omega_a (\eta_0 -\chi ,\chi\hat{\bm n})
		=\int\frac{\dd^3{\bm k}}{(2\pi )^3}
			\sum_{m=-2}^2{}_{\pm 1}\Omega_{\bm k}^{(m)}(\eta_0 -\chi )\,
			{}_{\pm 1}G_{|m|}{}^m(\chi\hat{\bm n} ,\hat{\bm n},{\bm k})
	\,,\label{eq:_pm 1Omega expand}
}
where
\al{
	&\Upsilon_{\bm k}^{(0)}=\Psi_{\bm k}-\Phi_{\bm k}
	\,,\ \ \ 
	\Upsilon_{\bm k}^{(\pm 1)}
		=\sigma_{\rmg ,{\bm k}}^{(\pm 1)}
	\,,\ \ \ 
	\Upsilon_{\bm k}^{(\pm 2)}
		=\frac{1}{\sqrt{3}}h_{\bm k}^{(\pm 2)}
	\,,\\
	&{}_s\Omega_{\bm k}^{(0)}=0
	\,,\ \ \ 
	{}_s\Omega_{\bm k}^{(\pm 1)}
		=-\sqrt{2}\,s\,\sigma_{\rmg ,{\bm k}}^{(\pm 1)}
	\,,\ \ \ 
	{}_s\Omega_{\bm k}^{(\pm 2)}
		=-\sqrt{2}\,s\,h_{\bm k}^{(\pm 2)}
	\,,\ \ (\text{for}\ s=\pm 1)
}
To derive the explicit expression for the angular power spectrum
for the lensing potentials, we expand 
the gradient-/curl-modes by the basis functions as
\al{
	x=\int\frac{\dd^3{\bm k}}{(2\pi )^3}
			\sum_{L=1}^\infty\sum_{m=-2}^2\,
			\hat x_L{}^{(m)}\,{}_0G_L{}^m
	\,.
}
Substituting eqs.~\eqref{eq:TAM wave expand}\,, \eqref{eq:Theta expand}\,, and \eqref{eq:_pm 1Omega expand}
into eqs.~\eqref{eq:pspin mspin phi} and \eqref{eq:pspin mspin varpi}\,, and
using the properties of the spin-weighted spherical harmonics (see Appendix \ref{sec:Spherical harmonics}),
we obtain
\al{
	\frac{\hat\phi_\ell{}^{(m)}}{2\ell +1}
		=&\int^{\cS}_0\dd\chi\frac{\cS -\chi}{\cS\chi}
			\Biggl[\,
				\Upsilon_{\bm k}^{(m)}(\eta_0 -\chi )\,
				{}_0\epsilon_\ell^{(|m|,m)}(k\chi )
	\notag\\
	&\quad\quad
				+\sqrt{\frac{1}{\ell (\ell +1)}}
					\frac{\dd}{\dd\chi}
						\biggl\{
							\chi
							\Bigl(
								{}_{+1}\Omega_{\bm k}^{(m)}(\eta_0 -\chi )
							\Bigr)\,
							{}_1\epsilon_\ell^{(|m|,m)}(k\chi )
						\biggr\}
			\Biggr]
	\,,\\
	\frac{\hat\varpi_\ell{}^{(m)}}{2\ell +1}
		=&\sqrt{\frac{1}{\ell (\ell +1)}}
				\int^{\cS}_0\dd\chi\frac{\cS -\chi}{\cS\chi}
						\frac{\dd}{\dd\chi}
						\biggl\{
							\chi
							\Bigl(
								{}_{+1}\Omega_{\bm k}^{(m)}(\eta_0 -\chi )
							\Bigr)
							{}_1\beta_\ell^{(|m|,m)}(k\chi )
						\biggr\}
	\,.
}
These integral solutions can be rewritten with
\al{
	\frac{\hat x_\ell{}^{(m)}}{2\ell +1}
		=\int^{\cS}_0k\,\dd\chi\,\Psi_{\bm k}^{(m)}(\eta_0 -\chi )\,\mcS_{x,\ell}^{(m)}(k,\chi )
	\,,
}
where we have defined the useful quantities as
\al{
	\Psi_{\bm k}^{(0)}=\frac{1}{2}\left(\Psi_{\bm k} -\Phi_{\bm k}\right)
	\,,\ \ 
	\Psi_{\bm k}^{(\pm 1)}=\sigma_{\rmg ,{\bm k}}^{(\pm 1)}
	\,,\ \ 
	\Psi_{\bm k}^{(\pm 2)}=2\,h_{\bm k}^{(\pm 2)}
	\,.\label{eq:Psi_k^m}
}
The transfer functions $\mcS_{x,\ell}^{(m)}$ are given by
\al{
	&\mcS_{\phi ,\ell}^{(0)}
		=2\frac{\cS -\chi}{\cS}\frac{1}{k\chi}{}_0\epsilon_\ell^{(0,0)}(k\chi )
	\,,\label{eq:S_phi,ell^(0)}\\
	&\mcS_{\phi ,\ell}^{(\pm 1)}
		=\frac{1}{k\chi}
			\Biggl[
				\frac{\cS -\chi}{\cS}{}_0\epsilon_\ell^{(1,\pm 1)}(k\chi )
				-\sqrt{2\frac{(\ell -1)!}{(\ell +1)!}}\,
					{}_1\epsilon_\ell^{(1,\pm 1)}(k\chi )
			\Biggr]
	\,,\\
	&\mcS_{\phi ,\ell}^{(\pm 2)}
		=\frac{1}{2k\chi}
			\Biggl[
				\frac{\cS -\chi}{\sqrt{3}\cS}{}_0\epsilon_\ell^{(2,\pm 2)}(k\chi )
				-\sqrt{2\frac{(\ell -1)!}{(\ell +1)!}}\,
				{}_1\epsilon_\ell^{(2,\pm 2)}(k\chi )
			\Biggr]
			+\frac{1}{10\sqrt{3}}\delta_{\ell ,2}\delta_\rmD (k\chi )
	\,,\label{eq:S_phi,ell^(2)}
}
for the gradient-mode lensing potential, and
\al{
	\mcS_{\varpi ,\ell}^{(0)} &= 0
	\,, \\ 
	\mcS_{\varpi ,\ell}^{(\pm 1)}
		&= -\sqrt{2\frac{(\ell -1)!}{(\ell +1)!}}\,
			\frac{1}{k\chi}\,{}_1\beta_\ell^{(1,\pm 1)}(k\chi )
	\,,\label{eq:S_varpi,ell^(1)}\\
	\mcS_{\varpi ,\ell}^{(\pm 2)}
		&= -\sqrt{\frac{1}{2}\frac{(\ell -1)!}{(\ell +1)!}}\,
			\frac{1}{k\chi}\,{}_1\beta_\ell^{(2,\pm 2)}(k\chi )
	\,,\label{eq:S_varpi,ell^(2)}
}
for the curl-mode lensing potential.

The gauge-invariant quantities of metric perturbations
contain statistical information for spatial randomness.
The quantities, 
$\Psi^{(m)}_{\bm k}=((\Psi_{\bm k}-\Phi_{\bm k})/2\,,\sigma_{\rmg ,{\bm k}}^{(\pm 1)}\,,2h_{\bm k}^{(\pm 2)})$\,, 
are responsible for randomness arising from initial
condition and late time evolution.
Assuming the unpolarized state of the perturbations, we characterize their
statistical properties as
\al{
	&\ave{
		\Bigl(\Psi_{\bm k}^{(m)}(\eta_0-\chi )\Bigr)^*
		\Psi_{{\bm k}'}^{(m')}(\eta_0-\chi' )
	}
		=(2\pi )^3\delta_{mm'}\delta_\rmD^3 ({\bm k}-{\bm k}')
                \mcP_{|m|}(k;\chi ,\chi' )
	\,,\label{eq:P_PsiPsi def}
}
Because of statistical isotropy, the power spectrum of the coefficients, 
$\hat\phi_\ell{}^{(m)}({\bm k})$ and $\hat\varpi_\ell{}^{(m)}({\bm k})$\,, depends
only on $k$\,.
Hence, we introduce their angular power spectrum which is of the form:
\al{
	&\ave{\left(\hat x_\ell{}^{(m)}({\bm k})\right)^*\hat x'_\ell{}^{(m')}({\bm k}')}
	=(2\pi )^3\delta_{mm'}\delta_\rmD^3 ({\bm k}-{\bm k}')\,(2\ell +1)^2\,
                \mcM^{xx'}_\ell(k)
	\,,
}
where
\al{
	&\mcM^{xx}_\ell{}^{(m)}(k)
		=\int^{\cS}_0k\,\dd\chi\int^{\cS}_0k\,\dd\chi'\,
			\mcS_{x,\ell}^{(m)}(k,\chi )
			\mcS_{x,\ell}^{(m)}(k,\chi' )\,
			\mcP_{|m|}(k;\chi ,\chi' )
	\,.\label{eq:M^xx(m) def}
}
There is no cross correlation $\mcM^{\phi\varpi}_\ell{}^{(m)}=0$ 
due to the parity symmetry.
With these notations, the auto- and cross-angular power spectrum leads to
\al{
	C_\ell^{xx'}
		=\frac{2}{\pi}\int^\infty_0\dd k\,k^2
			\sum_{m=-2}^2\,\mcM_\ell^{xx'}{}^{(m)}(k)
	\,,\label{eq:C_l^xx}
}
where $x$ and $x'$ take on $\phi$ and $\varpi$\,.
The $m=0,\pm 1,\pm 2$ modes corresponds to the scalar, vector, and tensor metric perturbations.
One clearly sees that the curl-mode is not generated by
the $m=0$ mode (scalar metric perturbations),
but is sourced by vector and tensor modes, as is expected.
The resultant angular power spectrum induced by the $m=\pm 1$ modes
(vector metric perturbations) exactly coincides with those derived
from a different method (eqs.~(3.26)-(3.27) of \cite{Yamauchi:2012bc}).

\subsection{E- and B-modes}
\label{sec:E- and B-modes}

Let us consider the cosmic shear field, based on eq.~\eqref{eq:linear-order Jacobi map}\,.
In terms of the polarization basis \eqref{eq:polarization basis}\,,
the Jacobi map can be decomposed into the spin-$0$ and spin-$\pm 2$
components:
\al{
	&{}_0\tilde\mcD
		=\tilde\mcD_{ab}e^a_+e^b_-
	\,,\ \ \ 
	{}_{\pm 2}\tilde\mcD
		=\tilde\mcD_{ab}e^a_\pm e^b_\pm 
	\,.
}
In practice, our actual observable is the ellipticity of the galaxy image,
which is the ratio of the anisotropic to isotropic deformation.
This is described by the reduced shear $g$ and $g^*$ defined through
the spin-weighted Jacobi map:
\al{
	g=-\frac{{}_{+2}\tilde\mcD}{{}_0\tilde\mcD}
	\,,\ \ \ 
	g^*=-\frac{{}_{-2}\tilde\mcD}{{}_0\tilde\mcD}
	\,.
}
Since ${}_0\mcD\approx 2\cS$ at linear-order, 
the reduced shear is simply related to the shear field:
\al{
	&g \approx -\frac{{}_{+2}\tilde\mcD}{2\cS}
		=-\frac{1}{2}\gamma_{ab}e^a_+e^b_+
	\,, \\ 
	&g^* \approx -\frac{{}_{-2}\tilde\mcD}{2\cS}
		=-\frac{1}{2}\gamma_{ab}e^a_-e^b_-
	\,.\label{eq:reduced shear}
}
Since the reduced shear is transformed as the spin-$\pm 2$ quantities, they
are decomposed by the spin-$\pm 2$ spherical harmonics as 
\al{
	&g = \sum_{\ell =2}^\infty\sum_{m=-\ell}^\ell\,
			\bigl( E_{\ell m}+i\,B_{\ell m}\bigr)\,
			{}_{+2}Y_{\ell m}
	\,, \\ 
	&g^* = \sum_{\ell =2}^\infty\sum_{m=-\ell}^\ell\,
			\bigl( E_{\ell m}-i\,B_{\ell m}\bigr)\,
			{}_{-2}Y_{\ell m}
	\,.\label{eq:reduced shear Y_lm expansion}
}
Here, $E_{\ell m}$ and $B_{\ell m}$ represent the two
parity eigenstate with electric-type $(-1)^\ell$
and magnetic-type $(-1)^{\ell +1}$ parities, respectively.
We then define the auto- and cross-angular power spectrum of these quantities as
\al{
	C_\ell^{\rmX\rmX'}
		=\frac{1}{2\ell +1}\sum_{m=-\ell}^\ell\,\ave{\left( X_{\ell m}\right)^* X'_{\ell m}}
	\,,\label{eq:C_l^XX' def}
}
where $X\,,X'=E\,,B$\,.
In terms of the spin-raising and lowering operators (see Appendix \ref{sec:Intrinsic covariant derivative}), 
the reduced shear \eqref{eq:reduced shear} is rewritten as
\al{
	&g=-\frac{1}{2}
				\int^{\cS}_0\dd\chi\frac{\cS -\chi}{\cS\chi}
				\Biggl\{
					\pspin^2\,\Upsilon
					+\frac{\dd}{\dd\chi}
						\Bigl(
							\chi\,\pspin\bigl( e^a_+\Omega_a\bigr)
						\Bigr)
				\Biggr\}
				-\frac{1}{4}\bigl[e^a_+e^b_+h_{ab}\bigr]^{\cS}_0
	\,,\label{eq:g} \\
	&g^*=-\frac{1}{2}
				\int^{\cS}_0\dd\chi\frac{\cS -\chi}{\cS\chi}
				\Biggl\{
					\mspin^2\,\Upsilon
					+\frac{\dd}{\dd\chi}
						\Bigl(
							\chi\,\mspin\bigl( e^a_-\Omega_a\bigr)
						\Bigr)
				\Biggr\}
				-\frac{1}{4}\bigl[e^a_-e^b_-h_{ab}\bigr]^{\cS}_0
	\,,\label{eq:g^*}
}
where we have used the relations:
${}_0X_{:ab}e^a_+e^b_+=\pspin^2\left({}_0X\right)$\,,
${}_0X_{:ab}e^a_-e^b_-=\mspin^2\left({}_0X\right)$\,,
$X_{a:b}e^a_+e^b_+=-\pspin\left( e^a_+X_a\right)$\,,
and $X_{a:b}e^a_-e^b_-=-\mspin\left( e^a_-X_a\right)$\,.
In order to derive the explicit expression for the angular power spectrum for
the E-/B-mode shear, we expand the reduced shear by
the basis functions as
\al{
	&g=-\int\frac{\dd^3{\bm k}}{(2\pi )^3}
				\sum_{L=2}^\infty\sum_{m=-2}^2\,
				\Bigl( 
					\hat E_L{}^{(m)}+i\hat B_L{}^{(m)}
				\Bigr)\,
				{}_{+2}G_L{}^m
	\,, \\
	&g^*=-\int\frac{\dd^3{\bm k}}{(2\pi )^3}
				\sum_{L=2}^\infty\sum_{m=-2}^2\,
				\Bigl( 
					\hat E_L{}^{(m)}-i\hat B_L{}^{(m)}
				\Bigr)\,
				{}_{-2}G_L{}^m
}
Expanding $e^a_\pm e^b_\pm h_{ab}$ in the same way as eqs.~\eqref{eq:Theta expand} and \eqref{eq:_pm 1Omega expand}\,,
\al{
	e^a_\pm e^b_\pm h_{ab}(\eta_0 -\chi ,\chi\hat{\bm n})
		=\int\frac{\dd^3{\bm k}}{(2\pi )^3}\sum_{m=\pm 2}2\sqrt{2}\,
			h_{\bm k}^{(m)}(\eta_0 -\chi )\,
			{}_{\pm 2}G_2{}^m(\chi\hat{\bm n},\hat{\bm n},{\bm k})
	\,,
}
substituting eqs.~\eqref{eq:TAM wave expand}\,, \eqref{eq:Theta expand}\,, and \eqref{eq:_pm 1Omega expand}
into eqs.~\eqref{eq:g} and \eqref{eq:g^*}\,, and using the properties of the spin-weighted spherical 
harmonics (see Appendix \ref{sec:Spherical harmonics})\,,
we obtain the integral solution for the E-/B-mode shear:
\al{
	\frac{\hat E_\ell{}^{(m)}}{2\ell +1}
		=&\frac{1}{2}\sqrt{\frac{(\ell +2)!}{(\ell -2)!}}
				\int^{\cS}_0\dd\chi\frac{\cS -\chi}{\cS\chi}
				\Biggl[
						\Upsilon_{\bm k}^{(m)}(\eta_0 -\chi )\,
						{}_0\epsilon_\ell^{(|m|,m)}(k\chi )
	\notag\\
	&\quad\quad\quad
					+\sqrt{\frac{1}{\ell (\ell +1)}}
						\frac{\dd}{\dd\chi}
							\biggl\{
								\chi
								\Bigl(
								{}_{+1}\Omega_{\bm k}^{(m)}(\eta_0 -\chi )
								\Bigr)
								{}_1\epsilon_\ell^{(|m|,m)}(k\chi )
							\biggr\}
				\Biggr]
	\notag\\
	&\quad\quad\quad
			+\frac{1}{\sqrt{2}}\delta_{m,\pm 2}
				\Bigl[
					h_{\bm k}^{(m)}(\eta_0 -\chi )
					{}_2\epsilon_\ell^{(2,m)}(k\chi )
				\Bigr]^{\cS}_0
	\,,\\
	\frac{\hat B_\ell{}^{(m)}}{2\ell +1}
		=&\frac{1}{2}\sqrt{\frac{(\ell +2)!(\ell -1)!}{(\ell -2)!(\ell +1)!}}
				\int^{\cS}_0\dd\chi\frac{\cS -\chi}{\cS\chi}
						\frac{\dd}{\dd\chi}
							\biggl\{
								\chi
								\Bigl(
									{}_{+1}\Omega_{\bm k}^{(m)}(\eta_0 -\chi )
								\Bigr)
								{}_1\beta_\ell^{(|m|,m)}(k\chi )
							\biggr\}
	\notag\\
	&\quad
			+\frac{1}{\sqrt{2}}
				\delta_{m,\pm 2}
					\Bigl[
						h_{\bm k}^{(m)}(\eta_0 -\chi )
						{}_2\beta_\ell^{(2,m)}(k\chi )
					\Bigr]^{\cS}_0
	\,.
}
To discuss the weak lensing measurement with the imaging survey, we assume
the redshift distribution of background sources, $N(\cS )\dd\cS$\,.
Taking account of the redshift distribution of background sources, namely galaxies, 
we can recast the formula for the $E$ and $B$ mode shear as
\al{
	\frac{\hat X_\ell{}^{(m)}}{2\ell +1}
		=\int^\infty_0 k\,\dd\chi\,\Psi_{\bm k}^{(m)}(\eta_0 -\chi )\,\mcS_{\rmX ,\ell}^{(m)}(k,\chi )
	\,,
}
where $\Psi_{\bm k}^{(m)}$ have been defined in eq.~\eqref{eq:Psi_k^m},
the explicit expressions for the transfer functions $\mcS_{\rmX ,\ell}^{(m)}$
are given by
\al{
	&\mcS_{\rmE ,\ell}^{(0)}
		=\sqrt{\frac{(\ell -2)!}{(\ell +2)!}}
			\frac{1}{k\chi}
			\int^\infty_\chi\dd\cS\frac{\cS -\chi}{\cS}\frac{N(\cS )}{N_\rmg}
			\,{}_0\epsilon_\ell^{(0,0)}(k\chi )
	\,, \label{eq:S_E,ell^(0)} 
	\\
	&\mcS_{\rmE ,\ell}^{(\pm 1)}
		=\frac{1}{2}\sqrt{\frac{(\ell -2)!}{(\ell +2)!}}
			\frac{1}{k\chi}
			\int^\infty_\chi\dd\cS\frac{N(\cS )}{N_\rmg}
			\Biggl[
				\frac{\cS -\chi}{\cS}{}_0\epsilon_\ell^{(1,\pm 1)}(k\chi )
				-\sqrt{2\frac{(\ell -1)!}{(\ell +1)!}}\,
					{}_1\epsilon_\ell^{(1,\pm 1)}(k\chi )
			\Biggr]
	\,, 
	\\
	&\mcS_{\rmE ,\ell}^{(\pm 2)}
		=\frac{1}{4}\sqrt{\frac{(\ell -2)!}{(\ell +2)!}}
			\frac{1}{k\chi}
			\int^\infty_\chi\dd\cS\frac{N(\cS )}{N_\rmg}
			\Biggl[
				\frac{1}{\sqrt{3}}\frac{\cS -\chi}{\cS}
				{}_0\epsilon_\ell^{(2,\pm 2)}(k\chi )
				-\sqrt{2\frac{(\ell -1)!}{(\ell +1)!}}\,
				{}_1\epsilon_\ell^{(2,\pm 2)}(k\chi )
			\Biggr]
	\notag \\ 
	&\quad\quad\quad\quad\quad\quad
			+\frac{1}{10\sqrt{2}}\delta_{\ell ,2}\delta_\rmD (k\chi )
			+\frac{1}{2\sqrt{2}}\frac{N(\chi )}{kN_\rmg}\,{}_2\epsilon_\ell^{(2,\pm 2)}(k\chi )\,
	\,, \label{eq:S_E,ell^(2)}
}
and
\al{
	&\mcS_{\rmB ,\ell}^{(0)}=0
	\\ 
	&\mcS_{\rmB ,\ell}^{(\pm 1)}
		=-\sqrt{\frac{1}{2}\frac{(\ell +2)!(\ell -1)!}{(\ell -2)!(\ell +1)!}}\,
			\frac{1}{k\chi}
			\int^\infty_\chi\dd\cS\frac{N(\cS )}{N_\rmg}\,
			{}_1\beta_\ell^{(1,\pm 1)}(k\chi )
	\,,\label{eq:S_B,ell^(1)} 
	\\
	&\mcS_{\rmB ,\ell}^{(\pm 2)}
		=-\frac{1}{2}\sqrt{\frac{1}{2}\frac{(\ell +2)!(\ell -1)!}{(\ell -2)!(\ell +1)!}}\,
				\frac{1}{k\chi}
				\int^\infty_\chi\dd\cS\frac{N(\cS )}{N_\rmg}
				{}_1\beta_\ell^{(2,\pm 2)}(k\chi )
				+\frac{1}{2\sqrt{2}}\frac{N(\chi )}{kN_\rmg}\,{}_2\beta_\ell^{(2,\pm 2)}(k\chi )
	\,, \label{eq:S_B,ell^(2)}
}
where the quantity $N_\rmg$ is the total number of galaxies,
defined by $N_\rmg =\int^\infty_0\dd\cS N(\cS )$\,.
One can confirm that the contributions coming from 
the vector metric perturbations ($m=\pm 1$ modes)
and the tensor metric perturbation ($m=\pm 2$ modes)
coincide with those with those derived in Refs.~\cite{Yamauchi:2012bc} and \cite{Schmidt:2012nw},
respectively.
Note that the last terms in eqs.~\eqref{eq:S_E,ell^(2)}\,, \eqref{eq:S_B,ell^(2)}
characterize the contribution from the perturbations at the source position,
which corresponds to the metric shear/FNC term~\cite{Dodelson:2003bv,Schmidt:2012ne,Schmidt:2012nw}.

Now, similar manner to the lensing potential of deflection angle, 
we derive the auto- and cross-angular
power spectrum for the E-/B-mode cosmic shear:
\al{
	C_\ell^{\rmX\rmX'}
		=\frac{2}{\pi}\int^\infty_0\dd k\,k^2
			\sum_{m=-2}^2\,\mcM_\ell^{\rmX\rmX'}{}^{(m)}(k)
	\,,\label{eq:C_l^XX}
}
where $X$ and $X'$ take on $E$ and $B$,  
\al{
	&\mcM^{\rmX\rmX}_\ell{}^{(m)}(k)
		=\int^\infty_0 k\,\dd\chi\int^\infty_0 k\,\dd\chi'\,
			\mcS_{\rmX ,\ell}^{(m)}(k,\chi )\mcS_{\rmX ,\ell}^{(m)}(k,\chi' )
			\mcP_{|m|}(k;\chi ,\chi' )
	\,.\label{eq:M^XX(m) def}
}
Note that there is no cross correlation: $\mcM_\ell^{\rmE\rmB}{}^{(m)}=0$ due to 
the parity symmetry.

Consequently, comparing eqs.~\eqref{eq:S_phi,ell^(0)}-\eqref{eq:C_l^xx} and
eqs.~\eqref{eq:S_E,ell^(0)}-\eqref{eq:M^XX(m) def}\,, we find that
 the simple relation between the angular power spectra
for the E-/B-mode cosmic shear and the gradient-/curl-mode lensing potential, 
$C_\ell^{\phi\phi}=4\frac{(\ell -2)!}{(\ell +2)!}C_\ell^{\rmE\rmE}$
and $C_\ell^{\varpi\varpi}=4\frac{(\ell -2)!}{(\ell +2)!}C_\ell^{\rmB\rmB}$\,,
as previously reported in \cite{Stebbins:1996wx,Yamauchi:2012bc}\,,
does not hold for general metric perturbations even if the source
distributions for shear fields are same as that for the lensing potential of 
deflection angle.

\section{Applications}
\label{sec:Applications}

In this section, we give several examples of the utility of the 
full-sky formulae. In subsection \ref{sec:WL_scalar}, we 
first consider the weak lensing by density 
perturbations, and show that the standard formulae for weak lensing power 
spectra is reproduced. Then, in subsection \ref{sec:WL_vector_tensor}, 
we consider the cosmic strings and primordial gravitational waves as 
intriguing examples for vector and tensor perturbations. 
Based on the formulae, we explicitly compute the power spectra, showing 
the non-vanishing signals for 
B-mode cosmic shear and curl-mode lensing potential.

\subsection{Weak lensing by density (scalar) perturbations}
\label{sec:WL_scalar}

The density (scalar) perturbations give 
a major contribution to the weak lensing experiment, and as a simple 
example of the utility of our full-sky formulae, we will give explicit 
expressions for the weak lensing power spectra.

In a standard $\Lambda$CDM universe, 
the anisotropic stress vanishes and the scalar metric 
perturbations $\Psi_{\bm k}$ and $-\Phi_{\bm k}$ are the same, namely 
$\Psi_{\bm k}=-\Phi_{\bm k}=\Psi_{\bm k}^{(0)}$\,.
As we mentioned in previous section, 
the scalar metric perturbations ($m=0$ mode) 
cannot produce the curl-mode and the B-mode shear, 
$\mcS_{\varpi ,\ell}^{(0)}=\mcS_{\rmB ,\ell}^{(0)}=0$\,. 
Hence, we consider the gradient-mode and the E-mode shear. 
The cosmological Poisson equation relates the Bardeen potential $\Phi_{\bm k}$ 
to the density perturbations $\delta_{\bm k}$\, and it gives
\al{
	\Phi_{\bm k}(\eta )
		=-\frac{3}{2}\frac{\Omega_\rmm H_0^2}{k^2}\frac{1}{a(\eta )}
\delta_{\bm k}(\eta )
	\,.
}
Then, using the fact that $_0\epsilon_\ell^{(0,0)}(x)=j_\ell(x)$, 
from eqs.~\eqref{eq:S_phi,ell^(0)}\,, \eqref{eq:M^xx(m) def}\,, 
and \eqref{eq:C_l^xx}\,, the power spectrum of the gradient-mode lensing potential 
induced by the density perturbations ($m=0$ mode) becomes
\al{
&	C_\ell^{\phi\phi}
		=\frac{2}{\pi}\int^\infty_0\frac{\dd k}{k^2}\,
		\int^{\cS}_0\dd\chi \int^{\cS}_0\dd\chi'
	\notag\\
	&\quad\quad\quad\times
		\left(
			\frac{3\Omega_\rmm H_0^2}{a(\eta_0 -\chi)}\frac{\cS -\chi}{\cS\,\chi}j_\ell(k\chi)
		\right)
		\left(
			\frac{3\Omega_\rmm H_0^2}{a(\eta_0 -\chi')}\frac{\cS -\chi'}{\cS\,\chi'}j_\ell(k\chi')
		\right)
		P_{\delta\delta}(k;\chi,\chi'),
\label{eq:Cell_phi_phi}
}
where the quantity $P_{\delta\delta}$ is the power spectrum of density 
perturbations. On the other hand, with a help of 
\eqref{eq:S_E,ell^(0)}\,, \eqref{eq:C_l^XX}\,, and \eqref{eq:M^XX(m) def}\,, 
the power spectrum of E-mode shear induced by the $m=0$ modes becomes 
\al{
	C_\ell^{\rm EE}
	=\frac{(\ell+2)!}{(\ell-2)!}\,
\frac{2}{\pi}\int^\infty_0\frac{\dd k}{k^2}\,
\int_0^\infty \dd \chi \int_0^\infty \dd \chi' \, W(\chi)W(\chi')\,
j_\ell (k\chi )j_\ell (k\chi')\,P_{\delta\delta}(k;\chi,\chi')
}
with the kernel function defined by
\al{
W(\chi)=\frac{3}{2}\frac{\Omega_\rmm H_0^2}{a(\eta_0 -\chi)}
\int^\infty_{\chi}\dd\cS\frac{\cS -\chi}{\cS\,\chi}\frac{N(\cS)}{N_{\rm g}}.
}
These are the standard formulae for the weak lensing power spectra 
(e.g. see \cite{Hu:2000ee}). Note that for a source distribution at a 
single redshift, i.e., 
$N(\chi)=N_{\rm g}\,\delta_{\rm D}(\chi-\cS)$, we obtain 
$C_\ell^{\phi\phi}=4\,\frac{(\ell -2)!}{(\ell +2)!}\,C_\ell^{\rmE\rmE}$ 
\cite{Stebbins:1996wx,Yamauchi:2012bc}\,. 
The above expressions are further simplified if we assume the linear 
evolution of the density power spectrum, or consider the 
Limber approximation (flat-sky approximation).

\subsection{Weak lensing by vector and tensor perturbations}
\label{sec:WL_vector_tensor}

\subsubsection{Primordial gravitational waves}
\label{sec:Primordial gravitational wave}

Primordial gravitational waves 
generated in the very early universe are the 
representative passive source for tensor perturbations. 
Late-time evolution of the primordial gravitational waves is 
described as $h_{\bm k}^{(\pm 2)}(\eta)=T_h(k,\eta )\,h_{\bm k}^{(\pm 2)}(0)$\,,
where $T_h(k,\eta )$ is the transfer function.
Here, we adopt the transfer function of the analytic form, 
$T_h(k,\eta )=3j_1(k\eta )/k\eta $\,.  
While this is valid only in the matter dominated epoch, 
we keep using it 
just for a qualitative understanding of the behavior of the power spectrum.

The power spectrum of the curl-mode lensing potential is given by 
the formulae \eqref{eq:C_l^xx} with eq. \eqref{eq:M^xx(m) def}.  
For the contribution of tensor perturbation,  
we consider $m=\pm2$. Substituting 
eq.\eqref{eq:S_varpi,ell^(2)} into the formula, we obtain
(see also \cite{Dodelson:2003bv,Cooray:2005hm,Li:2006si,Schmidt:2012nw,Sarkar:2008ii}):
\al{
	C_\ell^{\varpi\varpi ,{\rm PGW(T)}}
		=4\pi\int^\infty_0\frac{\dd k}{k}\,
			\Delta_h^2 (k)
			\biggl[\,
				\frac{1}{2}\frac{(\ell -1)!}{(\ell +1)!}\sqrt{\frac{(\ell +2)!}{(\ell -2)!}}
				\int^{\cS}_0\dd\chi\frac{3j_1(k(\eta_0 -\chi ))}{k(\eta_0 -\chi )}\frac{j_\ell (k\chi )}{k\chi^2}
			\biggr]^2
	\,.
}
On the other hand, 
the B-mode power spectrum for primordial gravitational waves can be calculated 
from $m=\pm2$ of eq. \eqref{eq:C_l^XX} with \eqref{eq:M^XX(m) def}. 
Using eqs.~\eqref{eq:S_B,ell^(2)}\,, we have
\al{
	&C_\ell^{\rmB\rmB ,{\rm PGW(T)}}
		=4\pi\int^\infty_0\frac{\dd k}{k}\,
			\Delta_h^2 (k)
			\Biggl[\,
				\frac{1}{4}\int^\infty_0\dd\chi\frac{3j_1(k(\eta_0 -\chi ))}{k(\eta_0 -\chi )}
	\notag\\
	&\quad\times
					\Biggl\{
						\frac{(\ell +2)!(\ell -1)!}{(\ell -2)!(\ell +1)!}\,
						\int^\infty_\chi\dd\cS\frac{N(\cS )}{N_\rmg}\frac{j_\ell (k\chi )}{k\chi^2}
						-\frac{N(\chi )}{N_\rmg}\left( j_\ell' (k\chi )+2\frac{j_\ell (k\chi )}{k\chi}\right)
					\Biggr\}
			\Biggr]^2
	\,.
}
Here, the quantity $\Delta_h^2 (k)$ is the dimensionless 
primordial power spectrum, 
which is related to the power spectrum $\mcP_2$ of eq.~\eqref{eq:P_PsiPsi def}
through 
$(k^3/2\pi^2 )\mcP_2 (k,\chi ,\chi' )=\Delta_h^2 (k)T_h(k,\eta_0 -\chi )T_h(k,\eta_0 -\chi' )$\,.
We will assume a power-law spectrum characterized in the form 
$\Delta_h^2 (k)=r\Delta_\mcR^2 (k_{\rm pivot})(k/k_{\rm pivot})^{n_{\rm t}}$, where 
$\Delta_\mcR^2$ is the dimensionless power spectrum of primordial 
curvature perturbation. 

\subsubsection{A cosmic string network}
\label{sec:Cosmic string network}

A cosmic string network is an active seed that 
can continuously generate the metric fluctuations even at late-time epoch. 
Vector and tensor perturbations are sourced by the non-vanishing
stress-energy tensor. To be precise, these metric perturbations are 
related to the velocity perturbations $v_{\bm k}^{(\pm 1)}$ and
the anisotropic stress perturbations $\Pi_{\bm k}^{(\pm 2)}$ of the 
active seeds through~(e.g., \cite{Hu:1997hp})
\al{
	&\sigma_{\rmg ,{\bm k}}^{(\pm 1)}
		=\frac{16\pi Ga^2}{k^2}v_{\bm k}^{(\pm 1)}
	\,,\\
	&\ddot h_{\bm k}^{(\pm 2)}
		+2\frac{\dot a}{a}\dot h_{\bm k}^{(\pm 2)}+k^2h_{\bm k}^{(\pm 2)}
		=8\pi Ga^2\Pi_{\bm k}^{(\pm 2)}
	\,.\label{eq:tensor linearized Einstein eq}
}

Cosmic strings can appear naturally 
in the early universe through spontaneous symmetry breaking
(see e.g., \cite{Kibble:1976sj,Jeannerot:2003qv,Vilenkin-Shellard}) or
at the end of stringy inflation 
(see \cite{Sarangi:2002yt,Jones:2003da,Copeland:2003bj,Dvali:2003zj}). 
General properties of lensing by a cosmic string has been discussed 
in Refs.~\cite{Bernardeau:2000xu,Uzan:2000xv}.
Below, based on the velocity-dependent one-scale model 
(e.g., ~\cite{Martins:2000cs,Martins:1996jp}),  
we will explicitly compute the angular power spectrum for a cosmic 
string network. In this model, a string network is characterized by 
the correlation length $\xi =1/H\gamma_{\rm s}$\,, and the root-mean-square 
velocity $v_{\rm rms}$\,. Assuming the network approaches a scaling 
solution, the quantities $\gamma_{\rm s}$ and $v_{\rm rms}$ stay constant. 
Taking the probabilistic nature of the intercommuting process into 
account~\cite{Avgoustidis:2005nv,Takahashi:2008ui,Yamauchi:2010vy,Yamauchi:2010ms}, $\gamma_{\rm s}$ and $v_{\rm rms}$ are approximately described by
$\gamma_{\rm s}\approx (\pi\sqrt{2}/3\tilde cP)^{1/2}$ 
and $v_{\rm rms}^2\approx (1-\pi /3\gamma_{\rm s})/2$~\cite{Takahashi:2008ui}\,,
where $\tilde c\approx 0.23$ quantifies the efficiency of the 
loop formation~\cite{Martins:2000cs}, and $P$ is the intercommuting 
probability. In order to compute the weak lensing power spectra, we need 
to evaluate the correlation between the string segments, 
for which we adopt simple analytic model developed by 
\cite{Vincent:1996qr,Albrecht:1997mz,Hindmarsh:1993pu}.

For given expression of velocity and anisotropic stress power spectra, 
it is straightforward to compute the angular power spectra, 
but the derivation of the explicit expressions for those spectra involves several 
steps, which we present in appendix \ref{sec:Derivation of correlations 
of a cosmic string network} in detail (see also \cite{Yamauchi:2012bc}). 
Then, the power spectra of curl-mode lensing potential are obtained from 
the $m=\pm1$ and $\pm2$ terms of eq.~\eqref{eq:C_l^xx} with 
\eqref{eq:M^xx(m) def}. Substituting eqs.~\eqref{eq:S_varpi,ell^(1)} 
and \eqref{eq:S_varpi,ell^(2)} into the formula, the explicit expressions  
become
\al{
	&C_\ell^{\varpi\varpi ,{\rm CS(V)}}
		=4\pi\int^\infty_0\frac{\dd k}{k}
			\biggl[
	\sqrt{\frac{(\ell -1)!}{(\ell +1)!}}
		\int^{\cS}_0\frac{\dd\chi}{\chi}\Delta_1 (k,\chi)\, j_\ell (k\chi )
			\biggr]^2
	\,,\\
	&	C_\ell^{\varpi\varpi ,{\rm CS(T)}}
		=4\pi\int^\infty_0\frac{\dd k}{k}\,
			\biggl[\,
				\frac{1}{2}\frac{(\ell -1)!}{(\ell +1)!}\sqrt{\frac{(\ell +2)!}{(\ell -2)!}}
				\int^{\cS}_0\dd\chi\Delta_2 (k,\chi )\frac{j_\ell (k\chi )}{k\chi^2}
			\biggr]^2
	\,.
}
for vector and tensor perturbations, respectively. 
Similarly, for the B-mode shear,  
eq. \eqref{eq:C_l^XX}\,, with \eqref{eq:M^XX(m) def}, 
\eqref{eq:S_B,ell^(1)}\,and \eqref{eq:S_B,ell^(2)} leads to 
\al{
	&C_\ell^{\rmB\rmB ,{\rm CS(V)}}
		=4\pi\int^\infty_0\frac{\dd k}{k}
			\Biggl[
				\frac{1}{2}
				\sqrt{\frac{(\ell +2)!(\ell -1)!}{(\ell -2)!(\ell +1)!}}
				\int^\infty_0\frac{\dd\chi}{\chi}\int^\infty_\chi\dd\cS\frac{N(\cS )}{N_\rmg}
				\Delta_1 (k,\chi )j_\ell (k\chi )
			\Biggr]^2
	\,,\\
	&C_\ell^{\rmB\rmB ,{\rm CS(T)}}
		=4\pi\int^\infty_0\frac{\dd k}{k}\,
			\Biggl[\,
				\frac{1}{4}\int^\infty_0\dd\chi\Delta_2 (k,\chi )
	\notag\\
	&\quad\times
					\Biggl\{
						\frac{(\ell +2)!(\ell -1)!}{(\ell -2)!(\ell +1)!}\,
						\int^\infty_\chi\dd\cS\frac{N(\cS )}{N_\rmg}\frac{j_\ell (k\chi )}{k\chi^2}
						-\frac{N(\chi )}{N_\rmg}\left( j_\ell' (k\chi )+2\frac{j_\ell (k\chi )}{k\chi}\right)
					\Biggr\}
			\Biggr]^2
	\,. 
}
In the above, to evaluate the relevant integrals analytically, we assume
that the unequal-time auto-power spectrum defined by 
eq.~\eqref{eq:P_PsiPsi def} is separately evaluated as
\al{
	&\frac{k^3}{2\pi^2}\,\mcP_m (k;\chi ,\chi' )
		=\Delta_m (k,\chi )\Delta_m (k,\chi' )
	\,.
}
The $\Delta_m^2$ is the auto-power spectra induced by the cosmic strings, and 
is explicitly given by 
\al{
	&\Delta_1^2 (k,\chi)
	=\left( 16G\mu\right)^2
		\frac{\sqrt{6\pi}\, v_{\rm rms}^2}{12(1-v_{\rm rms}^2)}
			\frac{4\pi k^3\chi^2 a^4}{H}\left(\frac{a}{k\xi}\right)^5
					{\rm erf}\left(\frac{k\xi /a}{2\sqrt{6}}\right)
	\,,\label{eq:Delta_sigma_g}
}
for the vector metric perturbation, and
\al{
	&\Delta_2 (k,\chi)
		=\int^\eta k\,\dd\eta'\,\mcG (k\eta ,k\eta' )\,\mcK (k,\eta' )
	\,,
}
for the tensor perturbation. 
Here $\mcG$ is the Green function for eq.~\eqref{eq:tensor linearized Einstein eq}, and $\mcK$ is the kernel related to the anisotropic
stress perturbation $\Pi_{\bm k}^{(\pm 2)}$\,, given by
\al{
	&\mcK^2 (k,\eta )
		=\frac{\sqrt{6\pi}(8 G\mu )^2}{36\sqrt{1-v_{\rm rms}^2}}
		\frac{4\pi k^3\chi^2 a^4}{H}\left(\frac{a}{k\xi}\right)^5
		\Biggl[
			\Big\{ (1-v_{\rm rms}^2)^2+v^4_{\rm rms}\Bigr\}{\rm erf}\left(\frac{k\xi /a}{2\sqrt{6}}\right)
	\notag\\
	&\quad\quad\quad\quad\quad
			-\frac{6c_0^2(v_{\rm rms})}{(1-v_{\rm rms}^2)^2}\left(\frac{a}{k\xi}\right)^2
				\biggl\{
					{\rm erf}\left(\frac{k\xi /a}{2\sqrt{6}}\right)
					-\frac{k\xi /a}{\sqrt{6\pi}}e^{-\frac{1}{24}k^2\xi^2 /a^2}
				\biggr\}
		\Biggr]
	\,.\label{eq:K}
}
with the dimensionless string tension $G\mu$. 
The function 
${\rm erf}(x)=(2/\sqrt{\pi})\int^x_0\dd y e^{-y^2}$ is the error function.
In the scaling regime, $c_0$ can be approximately estimated as
$c_0(v_{\rm rms})\approx 
(2\sqrt{2}/\pi )v_{\rm rms}(1-v_{\rm rms}^2)(1-8v_{\rm rms}^6)/(1+8v_{\rm rms}^6)$\,.

\subsubsection{B-mode and curl-mode power spectra}
\label{sec:B-mode and curl-mode spectra}

\begin{figure}[tbp]
\begin{tabular}{cc}
\begin{minipage}{0.5\hsize}
\begin{center}
\includegraphics[width=\hsize ]{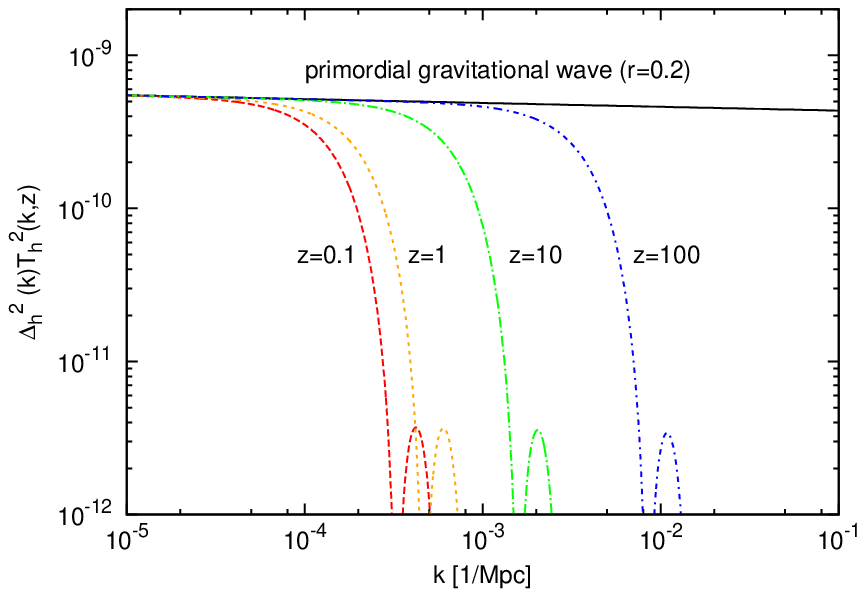}
\end{center}
\end{minipage}
\begin{minipage}{0.5\hsize}
\begin{center}
\includegraphics[width=\hsize ]{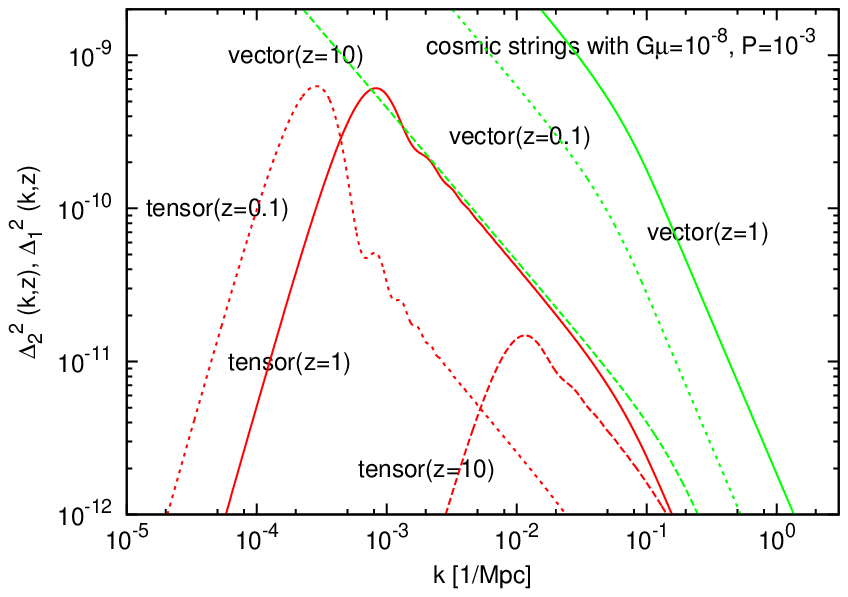}
\end{center}
\end{minipage}
\end{tabular}
\caption{ 
The dimensionless power spectrum for vector and tensor perturbations 
generated by primordial gravitational waves (left), a cosmic string network (right).
}
\label{Delta_h}
\end{figure}

Based on the formulae in sec.~\ref{sec:Cosmic string network} and 
\ref{sec:Primordial gravitational wave}, we now compute the weak lensing power spectra for 
the primordial gravitational waves and a cosmic string network.
In Fig.~\ref{Delta_h}, we first plot the dimensionless 
power spectra of the vector and tensor metric perturbations from 
the primordial gravitational waves (left) and 
a cosmic string network (right). 
Here, we specifically set the parameters 
to $G\mu =10^{-8}$ and $P=10^{-3}$ for the cosmic string network.
Qualitatively, 
all the power spectra are suppressed at small scales irrespective of the type 
of the metric perturbations. On the other hand, 
large-scale behaviors are rather different, and these are sensitive to 
the physical properties of the seeds. 
For the tensor perturbations, 
while the scale-invariant behavior of 
the spectrum of primordial gravitational waves merely reflects the 
initial condition, a negligible contribution of the large-scale 
fluctuations for the cosmic string network comes from the fact that 
the tensor fluctuation is produced by the motion of 
strings, and the typical scales of their fluctuations cannot exceed the size 
of cosmic strings. 
By contrast, the vector perturbation 
is directly related to the velocity perturbations, whose coherent length 
is typically larger than the size of seeds. As a result, the spectrum
of the vector perturbation for cosmic strings has a larger power,  
which scales as $\Delta_1^2\propto k^{-1}$ 
at large scales, and it dominates other  
perturbations.

\bc
\begin{figure}[tbp]
\bc
\hspace*{-0.4cm}
\includegraphics[width=79mm]{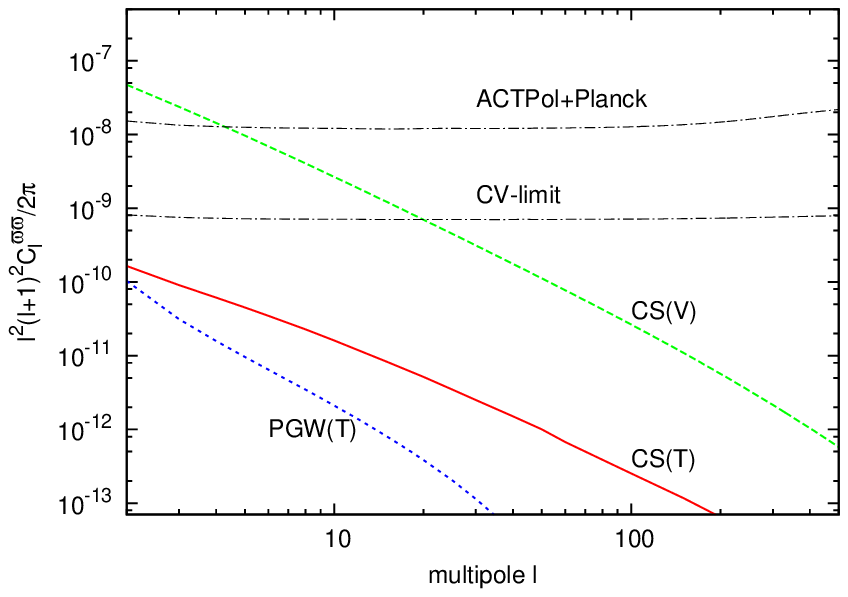}
\hspace*{-0.4cm}
\includegraphics[width=79mm]{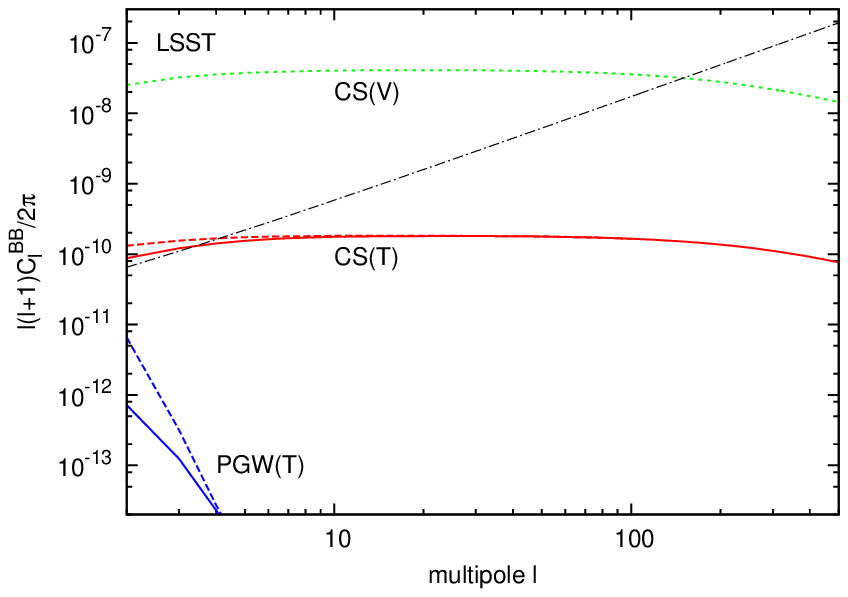}

\vspace*{0.0cm}

\caption{
{\it Left}: Angular power spectra of curl-mode lensing potential 
from the vector and tensor perturbations,   
generated by the primordial gravitational waves with $r=0.2$
(PGW(T); blue dotted), and the vector (CS(V); green dashed)
and the tensor (CS(T); red solid) contributions generated 
by the cosmic string network with $G\mu =10^{-8}$\,, $P=10^{-3}$.
The dot-dashed black lines represent the statistical errors 
coming from the reconstruction noise 
for the ACTPol+Planck and the CV-limit, with the maximum multipole
$\ell_{\rm max}=7000$~\cite{Namikawa:2011cs} used for reconstruction analysis.
{\it Right}: Angular power spectra of the B-mode cosmic shear
from the primordial gravitational wave (PGW(T); blue), 
and the tensor (CS(T); red) and vector (CS(V); green dotted) 
perturbations of the cosmic string network,
with (solid) and without (dashed) the metric shear/FNC term.
Here, we assume 
the LSST survey and adopt the redshift distribution of background galaxies,  
\eqref{eq:dndz_LSST}. 
The black dot-dashed line indicates the statistical error estimated from 
\eqref{eq:Delta_Cell_B}.
}
\label{varpi_B}
\ec
\end{figure} 
\ec

The distinctive features seen in Fig.~\ref{Delta_h} 
basically determine the shape and amplitude of the weak lensing power spectra. 
Fig.~\ref{varpi_B} shows the results of the angular power spectra for 
curl-mode lensing potential (left) and B-mode shear (right). 
The model parameters of the primordial gravitational waves and cosmic string 
network are the same as in Fig.~\ref{Delta_h}. The plotted angular 
power spectra are assumed to be measured from the CMB lensing experiment 
for the curl-mode lensing potential, 
and from the specific galaxy imaging survey, LSST, for the B-mode cosmic 
shear, respectively. That is, to compute the curl-mode power spectra, we set 
$\cS$ to the distance to the last scattering surface, 
while we adopt the following redshift distribution of background galaxies 
for the B-mode power spectra (e.g., \cite{arXiv:1103.1118,arXiv:1009.3204}): 
\al{
N(\cS )\dd\cS
=N_\rmg\frac{3z_\rmS^2}{(0.64z_\rmm )^3}
\exp\biggl[-\left(\frac{z_\rmS}{0.64z_\rmm}\right)^{3/2}\biggr]\dd z_\rmS
	\,,\label{eq:dndz_LSST}
}
with $z_\rmm=1.5$ and $N_\rmg =100\,[{\rm arcmin}^{-2}]$\,.

As it is expected, the primordial gravitational waves give a negligible 
contribution to the lensing power spectra, and 
it has only a small power at lower multipoles. This is fully consistent 
with previous works~\cite{Dodelson:2003bv,Schmidt:2012nw,Cooray:2005hm,Li:2006si,Sarkar:2008ii}.
The tensor perturbations induced by 
the cosmic strings are also shown to be a minor component of lensing spectrum, 
and it turns out that 
the vector perturbations from the cosmic strings can give 
the most dominant contribution among others. Here, for prospects of 
future detectability, we plot the expected statistical errors 
depicted as dot-dashed lines. For the curl-mode power spectrum, we 
consider the combination of the ground- and space-based high-resolution
CMB measurements by ACTPol and Planck (ACTPol+Planck) 
as well as an idealistic 
full-sky experiment only limited by the cosmic variance (CV-limit), 
and calculate the statistical error arising from the lens reconstruction 
method (see e.g., Ref.~\cite{Namikawa:2011cs}), 
assuming the maximum multipole $\ell_{\rm max}=7000$ 
used for reconstruction analysis. The statistical error for 
the B-mode spectrum is estimated from 
\al{
	\Delta C_\ell^{\rmB\rmB}=\sqrt{\frac{2}{(2\ell +1)f_{\rm sky}}}
			\frac{\ave{\gamma_{\rm int}^2}}{3600N_\rmg (180/\pi )^2}
	\,.\label{eq:Delta_Cell_B}
}
Assuming the LSST survey, we set the sky coverage to $f_{\rm sky}=0.5$, and 
adopt  the empirically estimated value of the 
root-mean-square intrinsic ellipticity,  
$\ave{\gamma_{\rm int}^2}^{1/2}=0.3$~\cite{astro-ph/0107431}\,.
Comparison between these statistical errors and the predictions of 
weak lensing power spectra immediately follows that 
it is very difficult and challenging to detect 
primordial gravitational waves via the weak lensing measurement 
(see also Refs.~\cite{Namikawa:2011cs,Schmidt:2012nw}), while 
a network of cosmic strings is potentially detectable through the 
measurement of curl-/B-mode spectrum by the vector perturbations.  
Although the actual impact on the detectability 
needs further consideration, it is found that 
the future weak lensing measurements have window 
to constrain the model parameters $G\mu$ and $P$ even tighter than 
the CMB observations through 
the Gott-Kaiser-Stebbins effect \cite{Yamauchi:2012bc}. 
In this respect, the curl-mode and B-mode lensing spectra can be used
as an important probe to find cosmic string network, and is complementary 
to the small-scale CMB experiment. Our 
full-sky formulae would be helpful for further study to 
explore the possibility to detect other exotic sources and to check the 
systematics.

\section{Summary} \label{sec:Summary} 

In this paper, we present a complete set of weak lensing power spectra
by scalar, vector, and tensor metric perturbations. 
Applying the total angular-momentum method, originally 
developed in the theoretical studies of CMB, we systematically derive 
the full-sky formulae for weak lensing observables such as 
the deflection angle and cosmic shear. In usual treatment of 
gravitational lensing,
the symmetric trace-free part of the angular gradient of the deflection angle 
can be used as a proxy for the cosmic shear, but the relations between 
these variables are in general non-trivial in the presence of all types of
the metric perturbations. 
Solving the first-order geodesic and geodesic deviation equations 
arising from the scalar, vector, and tensor perturbations, we have obtained 
the explicit gauge-invariant relation 
between the angular gradient of the deflection
angle $\Delta_a$ and the cosmic shear $\gamma_{ab}$ 
[eq.~\eqref{eq:gamma Delta relation}].

Then, based on the total angular momentum method, 
we presented the systematic construction of
the full-sky formulae of 
angular power spectra for the gradient-/curl-mode lensing potential 
of deflection angle [eqs.~\eqref{eq:S_phi,ell^(0)}-\eqref{eq:C_l^xx}], and
the E-/B-mode cosmic shear [eqs.~\eqref{eq:S_E,ell^(0)}-\eqref{eq:M^XX(m) def}].
To give examples of the utility of the formulae, 
we have considered the weak lensing by density (scalar) perturbation, and 
have shown that the standard forumula for the weak lensing power spectra 
(i.e., gradient- and E-mode spectra) is immediately reproduced. Further, 
as illustrative examples for vector and tensor perturbations, 
we have considered the primordial gravitational waves and 
cosmic string network.
Based on the formulae, 
we explicitly computed the power spectra, showing 
the non-vanishing signals for 
B-mode cosmic shear and curl-mode lensing potential. 
As shown in Fig.~\ref{varpi_B}, 
the weak lensing signal by vector perturbation from cosmic strings 
dominates other lensing signals, and  
is potentially detectable from future lensing experiments 
for a small intercommuting probability $P\ll1$.  
The framework presented here would thus be useful and helpful to 
explore the possibility to detect specific models for seeding
non-scalar metric perturbations.


\acknowledgments
This work is supported in part by a Grant-in-Aid for Scientific Research from 
the JSPS (No.~24540257). 
D.Y. is supported by Grant-in-Aid for JSPS Fellows (No.259800).


\appendix

\section{Useful formula} \label{sec:Formula}

In this appendix, we list some identities involving spherical Bessel function,
spin-weighted spherical harmonics, intrinsic covariant derivative, and
spin-operators which we have used in our calculations.

\subsection{Spherical Bessel function}
\label{sec:Spherical Bessel function}

The spherical Bessel functions, $j_\ell (x)$\,, are solutions to the differential equation:
\al{
	j_\ell'' (x)+\frac{2}{x}j_\ell' (x)
		+\left( 1-\frac{\ell (\ell +1)}{x^2}\right) j_\ell (x)=0
	\,.
}
The recursion relations of spherical Bessel functions are given by
\al{
	&\frac{j_\ell (x)}{x}
		=\frac{1}{2\ell +1}
			\Bigl( j_{\ell -1}(x)+j_{\ell +1}(x)\Bigr)
	\,,\ \ 
	j_\ell' (x)
		=\frac{1}{2\ell +1}
			\Bigl(\ell j_\ell (x)-(\ell +1)j_{\ell +1}(x)\Bigr)
	\,.
}

\subsection{Spin-weighted spherical harmonics}
\label{sec:Spherical harmonics}

To derive the spin-weighted spherical harmonics,
we first introduce the spin-weighted quantities and 
the spin-raising/lowering operators.
For a spin-$s$ function, ${}_sX$\,, we write in terms of the spin basis
and a symmetric trace-free rank-$s$ tensor, $X_{a_1\cdots a_s}$\,, as
\al{
	&{}_sX=X_{a_1\cdots a_s}e^{a_1}_+\cdots e^{a_s}_+
	\ \ (s\geq 0)
	\,,\ \ \ 
	{}_sX=X_{a_1\cdots a_{|s|}}e^{a_1}_-\cdots e^{a_{|s|}}_-
	\ \ (s<0)
	\,.
}
Furthermore, we define a pair of operator $\pspin$ and $\mspin$, called spin-raising
and lowering operators, respectively.
These operators have the properties of increasing or decreasing
the index of the spins by $1$\,.
For a spin-$s$ function ${}_sX$, these operators are defined as
\al{
	&\pspin\left({}_sX\right)
		\equiv -\sin^s\theta\left(\pd_\theta +\frac{i}{\sin\theta}\pd_\varphi\right)
				\sin^{-s}\theta\left({}_sX\right)
	\,,\label{eq:pspin def}\\
	&\mspin\left({}_sX\right)
		\equiv -\sin^{-s}\theta\left(\pd_\theta -\frac{i}{\sin\theta}\pd_\varphi\right)
				\sin^s\theta\left({}_sX\right)
	\,.\label{eq:mspin def}
}

The spin-weighted spherical harmonics can be obtained from the spherical harmonics
by applying the spin-raising and lowering operators.
The spin-weighted spherical harmonics of spin weight $s=0$ are simply
the standard spherical harmonics; ${}_0Y_\ell{}^m =Y_\ell{}^m$\,.
The spin-weighted spherical harmonics, ${}_sY_{\ell m}$\,, can be defined
in terms of the spin-$0$ spherical harmonics, $Y_{\ell m}$\,, as
\al{	
	&{}_sY_\ell{}^m
		=\sqrt{\frac{(\ell -s)!}{(\ell +s)!}}\,\pspin^s\, Y_{\ell m}
	\ \ (0\leq s\leq\ell )\,,\\
	&{}_sY_\ell{}^m
		=\sqrt{\frac{(\ell +s)!}{(\ell -s)!}}(-1)^s\,\mspin^{-s}\, Y_{\ell m}
	\ \ (-\ell\leq s\leq 0)\,,
}
and ${}_sY_\ell{}^m=0$ for $\ell <|s|$\,.
Here we have introduced the spin-operators defined in eqs.~\eqref{eq:pspin def}, \eqref{eq:mspin def}\,.
One can show that
\al{
	&\pspin\,{}_sY_\ell{}^m
		=\sqrt{(\ell -s)(\ell +s+1)}\,{}_{s+1}Y_\ell{}^m
	\,,\ \ 
	\mspin\,{}_sY_\ell{}^m
		=-\sqrt{(\ell +s)(\ell -s+1)}\,{}_{s-1}Y_\ell{}^m
	\,.
}
Since the spin-$0$\,, $\pm 1$\,, $\pm 2$ harmonics are useful in this paper,
we give their explicit form in Tables \ref{fig:spherical harmonics1}
and \ref{fig:spherical harmonics2}\,.
The spin-weighted spherical harmonics satisfy the conjugate relation
$({}_sY_\ell{}^m)^*=(-1)^{m+s}{}_{-s}Y_\ell{}^{-m}$\,,
the parity relation 
${}_sY_\ell{}^m (\pi -\theta ,\phi -\pi )=(-1)^\ell {}_{-s}Y_\ell{}^m (\theta ,\phi )$\,,
the orthonormal relationship 
\al{
	\int\dd\Omega\,\bigl({}_sY_{\ell'}{}^{m'}\bigr)^*{}_sY_\ell{}^m
		=\delta_{\ell\ell'}\delta_{mm'}
	\,,
}
the completeness relation
\al{
	\sum_{\ell =1}^\infty\sum_{m=-\ell}^\ell\,
		\bigl({}_sY_\ell{}^m (\theta' ,\phi' )\bigr)^*\,
		{}_sY_\ell{}^m (\theta ,\phi )
		=\delta_\rmD (\cos\theta -\cos\theta' )\delta_\rmD (\phi -\phi' )
	\,,
}
the addition relation
\al{
	\sum_m
		\bigl({}_{s_1}Y_\ell{}^m (\theta' ,\phi' )\bigr)^*\,
		{}_{s_2}Y_\ell{}^m (\theta ,\phi )
		=\sqrt{\frac{2\ell +1}{4\pi}}{}_{s_2}Y_\ell{}^{-s_1}(\beta ,\alpha )e^{-is_2\gamma}
	\,,
}
where $(\alpha ,\beta ,\gamma )$ relates the rotation from $(\theta' ,\phi' )$
through the origin to $(\theta ,\phi )$\,,
the Clebsch-Gordan relation
\al{
	{}_{s_1}Y_{\ell_1}{}^{m_1}\,{}_{s_2}Y_{\ell_2}{}^{m_2}
		=&\frac{\sqrt{(2\ell_1 +1)(2\ell_2 +1)}}{4\pi}
			\sum_{\ell ,m,s}
				\langle\ell_1 ,\ell_2 ;m_1,m_2|\ell_1 ,\ell_2 ;\ell ,m\rangle	
	\notag\\
	&\quad\quad\quad\times
				\langle \ell_1 ,\ell_2 ;-s_1,-s_2|\ell_1 ,\ell_2 ;\ell ,-s\rangle
				\sqrt{\frac{4\pi}{2\ell +1}}{}_sY_\ell{}^m
	\,,\label{eq:Clebsch-Gordan relation}
}
where $\langle\ell_1 ,\ell_2 ;m_1,m_2|\ell ,m\rangle$ denotes the Clebsch-Gordan
coefficient.

\bc
\begin{table}[t]
\bc
\caption{
Spin-$0$, $1$, $2$ spherical harmonics for $\ell =1$\,.
}
\vs{0.5}
\begin{tabular}{cccc} \hline \hline 
$m$ & ${}_0Y_1{}^m$ & ${}_{+1}Y_1{}^m$ & ${}_{+2}Y_1{}^m$\\ 
\hline
&&&\\
$0$&
$\sqrt{\frac{3}{4\pi}}\mu$&
$\sqrt{\frac{3}{8\pi}}\sqrt{1-\mu^2}$&
$0$\\
$\pm 1$&
$\mp\sqrt{\frac{3}{8\pi}}\sqrt{1-\mu^2}e^{\pm i\varphi}$&
$-\sqrt{\frac{3}{16\pi}}(1\mp\mu )e^{\pm i\varphi}$&
$0$\\
$\pm 2$&
$0$&
$0$&
$0$\\
&&&\\
\hline 
\end{tabular}
\label{fig:spherical harmonics1}
\ec
\end{table} 
\ec

\bc
\begin{table}[t]
\bc
\caption{
Spin-$0$, $1$, $2$ spherical harmonics for $\ell =2$\,.
}
\vs{0.5}
\begin{tabular}{cccc} \hline \hline 
$m$ & ${}_0Y_2{}^m$ & ${}_{+1}Y_2{}^m$ & ${}_{+2}Y_2{}^m$\\ 
\hline
&&&\\
$0$&
$\sqrt{\frac{5}{16\pi}}(3\mu^2 -1)$& 
$\sqrt{\frac{15}{8\pi}}\mu\sqrt{1-\mu^2}$&
$\sqrt{\frac{15}{32\pi}}(1-\mu^2 )$\\
$\pm 1$&
$\mp\sqrt{\frac{15}{8\pi}}\mu\sqrt{1-\mu^2}e^{\pm i\varphi}$&
$-\sqrt{\frac{5}{16\pi}}(1\mp\mu )(2\mu\pm 1)e^{\pm i\varphi}$&
$\sqrt{\frac{5}{16\pi}}\sqrt{1-\mu^2}(1\mp\mu )e^{\pm i\varphi}$ \\
$\pm 2$& 
$\sqrt{\frac{15}{32\pi}}(1-\mu^2 )e^{\pm 2i\varphi}$& 
$-\sqrt{\frac{5}{16\pi}}\sqrt{1-\mu^2}\left(\mu\mp 1\right) e^{\pm 2i\varphi}$&
$\sqrt{\frac{5}{64\pi}}\left( 1\mp\mu\right)^2 e^{\pm 2i\varphi}$ \\
&&&\\
\hline 
\end{tabular}
\label{fig:spherical harmonics2}
\ec
\end{table} 
\ec

\subsection{Intrinsic covariant derivative and spin operators}
\label{sec:Intrinsic covariant derivative}

In this subsection, we describe the basis vectors 
in the explicit Cartesian coordinate
and define the spin-raising/lowering operators, and present
the relation between the intrinsic covariant derivative on
the unit sphere and the spin operators.
When we consider a static observer, $u^\mu =(1,{\bm 0})$\,, 
the three-dimensional spatial basis vectors\,, $\hat{\bm n}$ and ${\bm e}_a$\,,
can be described explicitly in terms of the Cartesian coordinate as
\al{
	&\hat{\bm n}
		=\left(\sin\theta\cos\varphi ,\sin\theta\sin\varphi ,\cos\theta\right)
	\,,\\
	&{\bm e}_\theta (\hat{\bm n})
		=\left(\cos\theta\cos\varphi ,\cos\theta\sin\varphi ,-\sin\theta\right)
	\,,\\
	&{\bm e}_\varphi (\hat{\bm n})
		=\left( -\sin\theta\sin\varphi ,\sin\theta\cos\varphi ,0\right)
	\,.
}
With these notations, we have 
\al{
	 \hat n^i\pd_i =\pd_\chi
	\,,\ \ 
	 e^i_\theta\pd_i =\chi^{-1}\pd_\theta
	\,,\ \ 
	 e^i_\varphi =\chi^{-1}\pd_\varphi
	\,,\label{eq:derivative def}
}
and we can evaluate
\al{
	&\chi\,\hatn_{|j} e^j_a={\bm e}_a
	\,,\ \ 
	\chi^2\,\hatn_{|jk} e^j_{(a} e^k_{b)}=-\omega_{ab}\,\hatn
	\,,\ \ 
	{\bm e}_{a|j}\hat n^j =0
	\,,\label{eq:derivative relation 1}\\
	&\chi\,{\bm e}_{\theta |j} e^j_\varphi
		=\chi{\bm e}_{\varphi |j} e^j_\theta =\cot\theta\,{\bm e}_\varphi
	\,,\ \ 
	\chi\,{\bm e}_{\theta |j} e^j_\theta =-\hatn
	\,,\label{eq:derivative relation 2}\\
	&\chi\,{\bm e}_{\varphi |j} e^j_\varphi
		=-\sin\theta\left(\sin\theta\,\hatn +\cos\theta\,{\bm e}_\theta\right)
	\,.\label{eq:derivative relation 3}
}
We then derive the explicit relation between
the covariant derivative of a two-vector on the unit sphere
and the three-dimensional covariant derivative:
\al{
	X_{a:b}=\chi (X_ie^i_a)_{|j}e^j_b-{}^{(2)}\Gamma^c_{ab}X_ie^i_c
	\,,\ \ \ 
	{}^{(2)}\Gamma^c_{ab}=\chi e^i_{a|j}e^j_be^c_i
	\,,\label{eq:intrinsic covariant derivative}
}
where ${}^{(2)}\Gamma^c_{ab}$ denotes the Christoffel symbol defined on
the unit sphere.
Here the polarization indices are raised or lowered with respect to $\omega_{ab}$\,.
With a help of eqs.~\eqref{eq:derivative relation 2} and \eqref{eq:derivative relation 3}\,,
one can easily verify the following relations:
\al{
	e_\pm^a{}_{:b}\,e_\pm^b
		=\cot\theta\, e_\pm^a
	\,,\ \ \ 
	e_\pm^a{}_{:b}\,e_\mp^b
		=-\cot\theta\, e_\pm^a
	\,.\label{eq:derivative relation 4}
}
These equations can be used to construct the explicit relations
between the intrinsic covariant derivative and spin-raising/lowering operators.
Using eqs.~\eqref{eq:derivative relation 4}, one can verify the explicit relations such as 
${}_0X_{:ab}e^a_+e^b_-=\pspin\mspin\left({}_0X\right)$\,,
$X_{a:b}e^a_+e^b_+=-\pspin\left({}_{+1}X\right)$ and so on,
which we have used in our calculations.

\section{Christoffel symbols and Riemann tensors}
\label{sec:Christoffel symbols and Riemann tensors}

Christoffel symbols and Riemann tensor on the unperturbed spacetime in the Cartesian coordinate system 
are trivially $\Gamma^\rho_{\mu\nu}=0$\,, and $R_{\mu\rho\nu\sigma}=0$\,.
Since the Christoffel symbols are not covariant quantities, the unperturbed Christoffel
symbols in the spherical coordinate system can have the non-vanishing components:
\al{	
	&\Gamma^\chi_{ab}=-\chi\omega_{ab}
	\,,\ 
	\Gamma^a_{\chi b}=\frac{1}{\chi}\delta^a{}_b
	\,,\ 
	\Gamma^a_{bc}=^{(2)}\Gamma^a_{bc}
	\,,\ 
	\text{otherwise}=0
	\,,
}
where $a,b,c=\theta ,\varphi$ and the two-dimensional Christoffel symbols are given by
\al{	
	^{(2)}\Gamma^\theta_{\varphi\varphi}
		=-\sin\theta\cos\theta
	\,,\ \ \ 
	^{(2)}\Gamma^\varphi_{\theta\varphi}
		=\cot\theta
	\,,\ \ \ 
	\text{otherwise}=0
	\,.
}
Using the explicit expression for the linearized Christoffel symbols,
$\delta\Gamma^\rho_{\mu\nu}=\frac{1}{2}g^{\rho\sigma}
(\delta g_{\sigma\mu ;\nu}+\delta g_{\sigma\nu ;\mu}-\delta g_{\mu\nu ;\sigma})$\,,
we can calculate the components of the linearized Christoffel symbols as
\al{	
	&\delta\Gamma^0_{00}=\dot A
	\,,\ \ 
	\delta\Gamma^0_{0i}=A_{|i}
	\,,\ \ 
	\delta\Gamma^i_{00}=A^{|i}+\dot{\tilde B}^i
	\,,\ \ 
	\delta\Gamma^0_{ij}=-\tilde B_{(i|j)}+\frac{1}{2}\dot{\tilde H}_{ij}
	\,,\\
	&\delta\Gamma^i_{0j}
		=\frac{1}{2}\left( \tilde B^i{}_{|j}-\tilde B_j{}^{|i}\right) +\frac{1}{2}\dot{\tilde H}^i{}_j
	\,,\ \ 
	\delta\Gamma^i_{jk}
		=\frac{1}{2}\bar\gamma^{il}
			\left( \tilde H_{jl|k}+\tilde H_{kl|j}-\tilde H_{jk|l}\right)
	\,,
}
where we have introduced $\tilde B_i\equiv\delta g_{0i}$ and $\tilde H_{ij}\equiv\delta g_{ij}$
for simplicity.
With a help of the geodesic in the unperturbed spacetime, $x^\mu =(\eta_0 -\chi ,\chi\hat{\bm n})$\,,
and using the formulas \eqref{eq:derivative def}-\eqref{eq:derivative relation 3}\,,
we obtain the angular component of the linearized Christoffel symbols 
in terms of the gauge-invariant variables defined
in eqs.~\eqref{eq:delta g_ij def}-\eqref{eq:sigma_g,i def} as
\al{	
	 e_i^a\delta\Gamma^i_{\mu\nu}\frac{\dd x^\mu}{\dd\chi}\frac{\dd x^\nu}{\dd\chi}
		=&\frac{1}{\chi}\omega^{ab}
			\biggl\{
				\Upsilon_{:b}
				+\frac{\dd}{\dd\chi}
					\bigl(\chi\Omega_b\bigr)
			\biggr\}
	\,,
}
where we have introduced the spin-$0$ and $1$ combinations of the gauge-invariants: 
\al{
	&\Upsilon
		\equiv\Psi -\Phi +\sigma_{\rmg ,i}\hat n^i+\frac{1}{2}h_{ij}\hat n^i\hat n^j
	\,,\ \ 
	\Omega_i
		\equiv\sigma_{\rmg ,i}+h_{ij}\hat n^j
	\,.\label{eq:Upsilon Omega_i}
}
We can calculate the linearized Riemann tensor as 
$\delta R_{\mu\rho\nu\sigma}
=\frac{1}{2}(-\delta g_{\mu\nu ;\rho\sigma}-\delta g_{\rho\sigma ;\mu\nu}+\delta g_{\mu\sigma ;\rho\nu}+\delta g_{\nu\rho ;\mu\sigma})$\,.
We have the non-vanishing components of the linearized Riemann tensor as
\al{	
	&\delta R_{0i0j}
		=-A_{|ij}+\tilde B_{(i|j)}-\frac{1}{2}\ddot{\tilde H}_{ij}
	\,,\ \ 
	\delta R_{ij0k}
		=-\tilde B_{[i|j]k}-\dot{\tilde H}_{k[i|j]}
	\,,\\
	&\delta R_{ikj\ell}
		=\frac{1}{2}
			\left( 
				-\tilde H_{ij|k\ell}-\tilde H_{k\ell |ij}
				+\tilde H_{i\ell |kj}+\tilde H_{jk|i\ell}
			\right)
	\,.
}
Substituting these non-vanishing Riemann tensor into eq.~\eqref{eq:geodesic deviation equation}\,,
and using the relations between the basis vectors \eqref{eq:derivative def}-\eqref{eq:derivative relation 3}\,,
we have the explicit expression for the perturbed symmetric optical tidal matrix, $\delta\mcT_{ab}$\,, 
induced by the scalar, vector, and tensor perturbations:
\al{	
	\chi^2\delta\mcT_{ab}
		=&\Upsilon_{:ab}
			-\frac{\dd}{\dd\chi}\bigl(\chi\Omega_{(a:b)}\bigr)
			+\frac{1}{2}\chi\frac{\dd^2}{\dd\chi^2}\bigl(\chi h_{ab}\bigr)
		+\chi\omega_{ab}
			\biggl\{
				\pd_\chi\Upsilon -\frac{\dd}{\dd\chi}\left(\Omega_i\hat n^i\right)
				+\chi\frac{\dd^2}{\dd\chi^2}\mcR
			\biggr\}
	\,,\label{eq:chi^2 delta T_ab}
}
where $h_{ab}\equiv h_{ij}e^i_ae^j_b$\,, $\Omega_a\equiv\Omega_i e^i_a$\,, and
we have defined $\Upsilon$\,, $\Omega_i$ in eq.~\eqref{eq:Upsilon Omega_i}\,. 
We should note that the gauge degree of freedom is not completely removed
in the symmetric optical tidal matrix \eqref{eq:chi^2 delta T_ab}\,.
At first-order, however, the contributions from the residual gauge
freedom affects only the trace part of the symmetric optical
tidal matrix.
Hence, it is necessary to consider the other contributions 
when we take the trace part of the Jacobi map into account.

\section{${}_s\epsilon_L^{(\ell ,m)}$ and ${}_s\beta_L^{(\ell ,m)}$} 
\label{sec:Eplicit expression}

In this section we will present the explicit expression for 
${}_s\epsilon_L^{(\ell ,m)}$ and ${}_s\beta_L^{(\ell ,m)}$\,.
We are only interested in the cases of $s=0$\,, $\pm 1$\,, $\pm 2$\,.
Using the Clebsch-Gordan coefficients and the recurrent relations of
spherical Bessel functions presented in Appendix \ref{sec:Spherical Bessel function}\,, 
we obtain
\al{
	&{}_0\epsilon_L^{(0,0)}(x)
		=j_L (x)
	\,,\ \ \ 
	{}_0\epsilon_L^{(1,0)}(x)
		=j_L' (x)
	\,,\ \ \ 
	{}_0\epsilon_L^{(1,\pm 1)}(x)
		=\sqrt{\frac{L(L+1)}{2}}\frac{j_L(x)}{x}
	\,,\\
	&{}_0\epsilon_L^{(2,0)}(x)
		=\frac{1}{2}\left( 3j_L'' (x)+j_L (x)\right)
	\,,\ \ \ 
	{}_0\epsilon_L^{(2,\pm 1)}(x)
		=\sqrt{\frac{3L(L+1)}{2}}\left(\frac{j_L(x)}{x}\right)'
	\,,\\
	&{}_0\epsilon_L^{(2,\pm 2)}(x)
		=\sqrt{\frac{3(L+2)!}{8(L-2)!}}\frac{j_L(x)}{x^2}
	\,,\ \ \ 
	{}_0\beta_L^{(\ell ,m)}(x)=0\ \ (\ell =0\,,1\,,2)
	\,,
}
for $s=0$\,,
\al{
	&{}_1\epsilon_L^{(1,0)}(x)
		=\sqrt{\frac{L(L+1)}{2}}\frac{j_L(x)}{x}
	\,,\ \ \ 
	{}_1\epsilon_L^{(1,\pm 1)}(x)
		=\frac{1}{2}\left(\frac{j_L(x)}{x}+j_L'(x)\right)
	\,,\\
	&{}_1\epsilon_L^{(2,0)}(x)
		=\sqrt{\frac{3L(L+1)}{2}}\left(\frac{j_L(x)}{x}\right)'
	\,,\\
	&{}_1\epsilon_L^{(2,\pm 1)}(x)
		=\frac{1}{2}
			\left( j_L(x)+2j_L''(x)-2\frac{j_L(x)}{x^2}+\frac{2}{x}j_L'(x)\right)
	\,,\\
	&{}_1\epsilon_L^{(2,\pm 2)}(x)
		=\frac{1}{2}\sqrt{(L-1)(L+2)}
			\left(\frac{j_L(x)}{x^2}+\frac{j_L'(x)}{x}\right)
	\,,\\
	&{}_1\beta_L^{(1,0)}(x)=0
	\,,\ \ \ 
	{}_1\beta_L^{(1,\pm 1)}(x)
		=\pm\frac{1}{2}j_L(x)
	\,,\\
	&{}_1\beta_L^{(2,0)}(x)=0
	\,,\ \ \ 
	{}_1\beta_L^{(2,\pm 1)}(x)
		=\pm\frac{1}{2}\left( j_L'(x)-\frac{j_L(x)}{x}\right)
	\,,\\
	&{}_1\beta_L^{(2,\pm 2)}(x)
		=\pm\frac{1}{2}\sqrt{(L-1)(L+2)}\frac{j_L(x)}{x}
	\,,
}
for $s=\pm 1$\,, and
\al{
	&{}_2\epsilon_L^{(2,0)}(x)
		=\sqrt{\frac{3(L+2)!}{8(L-2)!}}\frac{j_L(x)}{x^2}
	\,,\\
	&{}_2\epsilon_L^{(2,\pm 1)}(x)
		=\frac{1}{2}\sqrt{(L-1)(L+2)}\left(\frac{j_L(x)}{x^2}+\frac{j_L'(x)}{x}\right)
	\,,\\
	&{}_2\epsilon_L^{(2,\pm 2)}(x)
		=\frac{1}{4}\left( -j_L(x)+j_L''(x)+2\frac{j_L(x)}{x^2}+4\frac{j_L'(x)}{x}\right)
	\,,\\
	&{}_2\beta_L^{(2,0)}(x)=0
	\,,\ \ \ 
	{}_2\beta_L^{(2,\pm 1)}(x)
		=\pm\frac{1}{2}\sqrt{(L-1)(L+2)}\frac{j_L(x)}{x}
	\,,\\
	&{}_2\beta_L^{(2,\pm 2)}(x)
		=\pm\frac{1}{2}\left( j_L'(x)+2\frac{j_L(x)}{x}\right)
	\,,
}
for $s=\pm 2$\,.

\section{Derivation of correlations of a cosmic string network} 
\label{sec:Derivation of correlations of a cosmic string network}

In this section, we derive the auto-power spectrum for the non-vanishing vector and
tensor modes of the string stress-energy tensor.
We first briefly review an analytic model for the correlation between
string segments in section \ref{sec:String correlators}\,.
In section \ref{sec:Vector mode} and \ref{sec:Tensor mode},
we derive the analytic expression for 
the auto-power spectrum for the vector and tensor perturbations 
induced by the cosmic string network.

\subsection{String correlators}
\label{sec:String correlators}

To evaluate the auto-power spectrum for the non-vanishing stress-energy tensor,
we write down the string stress-energy tensor.
The stress-energy tensor for a string segment in the transverse gauge is described as
\beq
	\delta T^{\mu\nu}({\bm r},\eta )
		=\mu\int\dd\sigma
			\left(
			\begin{array}{cc}
				1 & -\dot r^i\\
				-\dot r^j & \dot r^i\dot r^j -{r^i}'{r^j}'\\
			\end{array}
			\right)
			\delta_\rmD^3 ({\bm r}-{\bm r}(\sigma ,\eta ))
	\,,\label{eq:string stress-energy}
\eeq
where the dot ( $\dot{}$ ) and the prime ( ${}'$ ) denote 
the derivative with respect to $\eta$ and $\sigma$\,, respectively.
Hereafter, we use a simple model to compute the string correlations developed 
in Refs.~\cite{Vincent:1996qr,Albrecht:1997mz,Hindmarsh:1993pu}.
We assume that all correlations can be expressed in terms of 
the two point functions:
$\langle\dot r^i(\sigma_1 ,\eta )\dot r^j(\sigma_2 ,\eta )\rangle\,,\ 
\langle{r^i}'(\sigma_1 ,\eta ){r^j}'(\sigma_2 ,\eta )\rangle\,,\ 
\langle{r^i}'(\sigma_1 ,\eta )\dot r^j(\sigma_2 ,\eta )\rangle$\,.
Furthermore,these two point functions are assumed to be exactly Gaussian and isotropic
is assumed to be distributed with mean zero and following variances:
\al{
	&\big\langle\dot r^i(\sigma_1 ,\eta )\dot r^j(\sigma_2 ,\eta )\big\rangle
		=\frac{1}{3}\delta^{ij}\,V_{\rm s}(\sigma_1 -\sigma_2 ,\eta )
	\,,\label{eq:V_s def}\\
	&\big\langle{r^i}'(\sigma_1 ,\eta ){r^j}'(\sigma_2 ,\eta )\big\rangle
		=\frac{1}{3}\delta^{ij}\,T_{\rm s}(\sigma_1 -\sigma_2 ,\eta )
	\,,\label{eq:T_s def}\\
	&\big\langle{r^i}'(\sigma_1 ,\eta )\dot r^j(\sigma_2 ,\eta )\big\rangle
		=\frac{1}{3}\delta^{ij}\,M_{\rm s}(\sigma_1 -\sigma_2 ,\eta )
	\,.\label{eq:M_s def}
}
Since in our calculation we consider only a segment of a long string with length $\xi$\,,
the correlators are expected to be damped on scale larger than the correlation length, 
namely $\sigma\gg\xi /a$\,, and have the non-vanishing expectation value 
on $\sigma\ll\xi /a$. Hence the asymptotic forms are \cite{Hindmarsh:1993pu}
\al{
	&T_{\rm s} (\sigma ,\eta )=
		\Biggl\{
			\begin{array}{ll}
			\bar t^2\ &:\  \sigma\ll\xi /a\\
			0\  &:\  \sigma\gg\xi /a\\
			\end{array}
	\,,\label{eq:asymptotic behavior of T_s}\\
	&V_{\rm s} (\sigma ,\eta )=
		\Biggl\{
			\begin{array}{ll}
			\bar v^2\  &:\  \sigma\ll\xi /a\\
			0\  &:\  \sigma\gg\xi /a\\
			\end{array}
	\,,\\
	&M_{\rm s} (\sigma ,\eta )=
		\Biggl\{
			\begin{array}{ll}
			ac_0\sigma /\xi\  &:\  \sigma\ll\xi /a\\
			0\  &:\  \sigma\gg\xi /a\\
			\end{array}
	\,,
}
where we have introduced the three parameters:
$\bar t^2\equiv\big\langle{{\bm r}'}^2\big\rangle\,,\ 
\bar v^2\equiv\ave{\dot{\bm r}^2}\,,\ 
c_0\equiv\frac{\xi}{a}\ave{\dot{\bm r}\cdot{\bm r}''}$\,.
We can evaluate the parameters as
\al{
	&\bar t\approx\sqrt{1-v_{\rm rms}^2}
	\,,\ \ 
	\bar v\approx v_{\rm rms}
	\,,\ \ 
	c_0\approx\frac{2\sqrt{2}}{\pi}v_{\rm rms}\left( 1-v_{\rm rms}^2\right)\frac{1-8v_{\rm rms}^6}{1+8v_{\rm rms}^6}
	\,.
}
Once $\xi$ and $v_{\rm rms}$ are properly evaluated through the string network model,
they fix the parameters used to calculate the correlators.

\subsection{Vector mode}
\label{sec:Vector mode}

We provide brief summary of the method to calculate 
the auto-power spectrum for the vector perturbations in
the conformal Newton gauge, following \cite{Yamauchi:2012bc}.
The vector-type fluctuation for the stress-energy tensor can be decomposed into
the vector mode functions defined in section \ref{sec:Mode functions} as
\al{
	\delta T^0{}_i({\bm x},\eta )
		=\int\frac{\dd^3{\bm k}}{(2\pi )^3}
			\sum_{m=\pm 1}
				v_{\bm k}^{(m)}(\eta )\,Q^{(m)}_i({\bm x},{\bm k})
	\,.\label{eq:vector stress-energy}
}
Comparing to eqs.~\eqref{eq:vector stress-energy} and \eqref{eq:string stress-energy}\,,
we obtain the vector-type perturbations\,, $v_{\bm k}^{(\pm 1)}(\eta )$\,, due to a
string segment:
\al{
	v_{\bm k}^{(\pm 1)}(\eta )
		=\frac{\mu}{\sqrt{2}i}\int\dd\sigma\,
			{\bm e}_\pm^* (\hat{\bm k})\cdot\dot{\bm r}(\sigma ,\eta )
			e^{-i{\bm k}\cdot{\bm r}(\sigma ,\eta )}
	\,.
}
We assume that the total correlations can be described by a summation of the contribution
of each segment and the correlations between the different segments are negligibly small.
The linearized Einstein equation implies that the auto-power spectrum
for vector metric perturbations induced by the string network 
can be approximated as
\al{
	\Delta_1^2 (k,\eta )
		&=\frac{k^3}{2\pi^2}
			\frac{(16\pi G)^2a^4}{k^4}
			n_{\rm s}\delta V\frac{1}{\mcV}
			\ave{\left( v_{\bm k}^{(\pm 1)}(\eta )\right)^* v_{\bm k}^{(\pm 1)}(\eta )}
	\notag\\
	&
		=\frac{(16G)^2a^4}{4k}n_{\rm s}\delta V\frac{1}{\mcV}
			\int\dd\sigma_1\dd\sigma_2
			\left( e^{(\pm 1)}_i(\hat{\bm k})\right)^* e^{(\pm 1)}_j(\hat{\bm k})
	\notag\\
	&\quad\quad\quad\quad\times
			\ave{
				\dot r^i(\sigma_1 ,\eta )\dot r^j (\sigma_2 ,\eta )
				\exp
					\biggl[ 
						i{\bm k}\cdot\left({\bm r}(\sigma_1 ,\eta )-{\bm r}(\sigma_2 ,\eta )\right)
					\biggr]
				}
	\,,
}
where $\delta V=4\pi\chi^2 /H$ is the comoving differential volume element,
$n_{\rm s}=a^3\xi^{-3}$ is the comoving number density of the string segments, 
and $\mcV\equiv (2\pi )^3\delta_\rmD^3 ({\bm 0})$ is the comoving box size\,.
Using the correlators \eqref{eq:V_s def}-\eqref{eq:M_s def}\,,
we can compute the equal-time auto-power spectrum
for the tensor perturbations as
\beq
	\Delta_1^2 (k,\eta )
		=\frac{(16 G\mu )^2a^4}{24k}n_{\rm s}\delta V
			\frac{1}{\mcV}
			\int\dd\sigma_+\dd\sigma_- V_{\rm s}(\sigma_- ,\eta )
			\exp
				\biggl[
					-\frac{1}{6}k^2\Gamma_{\rm s}(\sigma_- ,\eta )
				\biggr]
	\,,
\eeq
where $\sigma_\pm =\sigma_1\pm\sigma_2$ and we have introduced $\Gamma_{\rm s}$ defined as
\al{
	&\Gamma_{\rm s}(\sigma_1 -\sigma_2 ,\eta )
		\equiv\ave{\Bigl\{{\bm r}(\sigma_1 ,\eta )-{\bm r}(\sigma_2 ,\eta )\Bigr\}^2}
		=\int^{\sigma_1}_{\sigma_2}\dd\sigma_3\dd\sigma_4\, T_{\rm s}(\sigma_3 -\sigma_4 ,\eta )
	\,.\label{eq:Gamma_s def}
}
The asymptotic behavior of $T_{\rm s}$, eq.~\eqref{eq:asymptotic behavior of T_s}\,,
leads to $\Gamma_{\rm s}\approx\bar t^2\sigma^2$ on scalar smaller than the correlation length.
Since the term $\int\dd\sigma_+ /\mcV$ corresponds to the length of the string
segment within the unit volume and the correlators are damped at $\sigma\gg\xi /a$\,, 
we can take the region of the integration as 
$\int\dd\sigma_+ /\mcV =a^2/\xi^2\sqrt{1-v_{\rm rms}^2}$
and $|\sigma_-|<\xi /2a\sqrt{1-v_{\rm rms}^2}$.
We then have
\beq
	\Delta_1^2 (k,\eta )
		=\left( 16G\mu\right)^2
					\frac{\sqrt{6\pi}\, v_{\rm rms}^2}{12(1-v_{\rm rms}^2)}
					\frac{4\pi k^3\chi^2 a^4}{H}\left(\frac{a}{k\xi}\right)^5
					{\rm erf}\left(\frac{k\xi /a}{2\sqrt{6}}\right)
	\,.\label{eq:equal-time auto-power spectrum}
\eeq

\subsection{Tensor mode}
\label{sec:Tensor mode}

For tensor-components, the fluctuation can be decomposed into
the tensor mode functions defined in section \ref{sec:Mode functions} as
\al{
	\delta T^i{}_j({\bm x},\eta )
		=\int\frac{\dd^3{\bm k}}{(2\pi )^3}
			\sum_{m=\pm 2}
				\Pi_{\bm k}^{(m)}(\eta )\,Q^{(m)}{}^i{}_j({\bm x},{\bm k})
	\,.\label{eq:tensor stress-energy}
}
Comparing to eqs.~\eqref{eq:string stress-energy} and \eqref{eq:tensor stress-energy}\,,
the tensor-type perturbations due to a string segment, $\Pi_{\bm k}^{(\pm 2)}(\eta )$\,,
are given by
\al{
	\Pi_{\bm k}^{(\pm 2)}(\eta )
		=&-\frac{1}{2\sqrt{2}}\,\mu
			\int\dd\sigma
				\left\{
					\dot r^i(\sigma ,\eta )\dot r^j(\sigma ,\eta )-{r^i}'(\sigma ,\eta ){r^j}'(\sigma ,\eta )
				\right\}
				e_{\pm ,i}^* (\hat{\bm k})e_{\pm ,j}^* (\hat{\bm k})\,
				e^{i{\bm k}\cdot{\bm r}(\sigma ,\eta )}
	\,.
}
Assuming that the auto-power spectra for the fluctuation of the stress-energy tensor
have the form:
\al{
	&\ave{
		\bigl(\Pi_{\bm k}^{(m)}(\eta )\bigr)^*\Pi_{{\bm k}'}^{(m')}(\eta' )
	}
		=(2\pi )^3\delta_\rmD^3 ({\bm k}-{\bm k}')\delta_{mm'}
			\frac{2\pi^2}{k^3}\Delta_\Pi (k,\eta )\Delta_\Pi (k,\eta' )
	\,,
}
we can describe the auto-power spectrum for the tensor perturbations
in terms of the auto-power spectrum for the tensor component
of the stress-energy as
\al{
	\Delta_2 (k,\eta )
		=\int^\eta k\,\dd\eta'\, \mcG (k\eta ,k\eta' )\,
			\mcK (k,\eta' )
	\,.
}
where $\mcG$ is the Green function for the equation \eqref{eq:tensor linearized Einstein eq} 
and $\mcK$ is the kernel defined by
\al{
	\mcK (k,\eta )
		\equiv\frac{8\pi Ga^2(\eta )}{k^2}\Delta_\Pi (k,\eta )
	\,.
}
With a help of the correlators eqs.~ \eqref{eq:V_s def}-\eqref{eq:M_s def} and \eqref{eq:Gamma_s def}\,, 
we can evaluate the auto-power spectrum for $\Pi$ as
\al{
	&\Delta_\Pi^2 (k,\eta )
		=\frac{k^3}{2\pi^2}
				n_{\rm s}\delta V\frac{1}{\mcV}
				\ave{{\Pi_{\bm k}^{(\pm 2)}}^*(\eta )\Pi_{\bm k}^{(\pm 2)}(\eta )}
	\notag\\
	&\quad
		=\frac{k^3\mu^2}{16\pi^2}n_{\rm s}\delta V\frac{1}{\mcV}\int\dd\sigma_1\dd\sigma_2\,
			e_{\pm ,i}^*(\hat{\bm k})e_{\pm ,j}^*(\hat{\bm k})
			e_{\pm ,k}(\hat{\bm k})e_{\pm ,\ell}(\hat{\bm k})\,
	\notag\\
	&\quad\quad\times
			\Big\langle
				\left\{
					\dot r^i(\sigma_1 ,\eta )\dot r^j(\sigma_1 ,\eta )-{r^i}'(\sigma_1 ,\eta ){r^j}'(\sigma_1 ,\eta )
				\right\}
				\left\{
					\dot r^k(\sigma_2 ,\eta )\dot r^\ell (\sigma_2 ,\eta )-{r^k}'(\sigma_2 ,\eta ){r^\ell}'(\sigma_2 ,\eta )
				\right\}
	\notag\\
	&\quad\quad\quad\quad\times
				\exp\Bigl[ 
					-i{\bm k}\cdot\Bigl({\bm r}(\sigma_1 ,\eta )-{\bm r}(\sigma_2 ,\eta )\Bigr)
					\Bigr]
			\Bigr\rangle
	\,.
}
With a help of the correlators defined in eqs.~\eqref{eq:V_s def}-\eqref{eq:M_s def}\,,
we can compute the equal-time auto-power spectrum
for the tensor-type perturbations as
\al{
	\Delta_\Pi^2 (k,\eta )
		=&\frac{k^3\mu^2}{36\pi^2}n_{\rm s}\delta V\frac{1}{\mcV}
			\int\dd\sigma_+\dd\sigma_-
	\notag\\
	&\quad\times
			\biggl(
				T_{\rm s}^2(\sigma_- ,\eta )+V_{\rm s}^2(\sigma_- ,\eta )-2M_{\rm s}^2(\sigma_- ,\eta )
			\biggr)
			\exp\left( -\frac{1}{6}k^2\Gamma (\sigma_- ,\eta )\right)
	\,,
}
where $\sigma_\pm\equiv\sigma_1\pm\sigma_2$ and we have defined 
$\Gamma_{\rm s}$ in eq.~\eqref{eq:Gamma_s def}\,.
As discussed in previous subsection \ref{sec:Vector mode}, 
taking the region of the integration as 
$\int\dd\sigma_+ /\mcV =a^2/\xi^2\sqrt{1-v_{\rm rms}^2}$
and $|\sigma_-|<\xi /2a\sqrt{1-v_{\rm rms}^2}$\,,
we have the kernel $\mcK$ induced by the string network:
\al{
	\mcK^2 (k,\eta )
		\approx &\frac{\sqrt{6\pi}(8 G\mu )^2}{36\sqrt{1-v_{\rm rms}^2}}
		\frac{4\pi k^3\chi^2 a^4}{H}\left(\frac{a}{k\xi}\right)^5
		\Biggl[
			\Big\{ (1-v_{\rm rms}^2)^2+v^4_{\rm rms}\Bigr\}{\rm erf}\left(\frac{k\xi /a}{2\sqrt{6}}\right)
	\notag\\
	&\quad\quad\quad
			-\frac{6c_0^2(v_{\rm rms})}{(1-v_{\rm rms}^2)^2}\left(\frac{a}{k\xi}\right)^2
				\biggl\{
					{\rm erf}\left(\frac{k\xi /a}{2\sqrt{6}}\right)
					-\frac{k\xi /a}{\sqrt{6\pi}}e^{-\frac{1}{24}k^2\xi^2 /a^2}
				\biggr\}
		\Biggr]
	\,.
}


\end{document}